\documentclass[12pt,preprint]{aastex}
\newcommand{\etal}{\mbox{\rm et al.}}

\newcommand{\lsun}{\mbox{$L_{\odot}$}}
\newcommand{\msun}{\mbox{$M_{\odot}$}}
\newcommand{\rsun}{\mbox{$R_{\odot}$}}
\newcommand{\lstar}{\mbox{$L_{\star}$}}
\newcommand{\mstar}{\mbox{$M_{\star}$}}
\newcommand{\rstar}{\mbox{$R_{\star}$}}

\newcommand{\bc}{\mbox{BC}}

\newcommand{\mv}{\mbox{$M_{\rm V}$}}

\newcommand{\vsini}{\mbox{$v \sin i$}}
\newcommand{\arel}{\mbox{$a_{\rm rel}$}}
\newcommand{\teff}{\mbox{$T_{\rm eff}$}}
\newcommand{\fe}{\mbox{\rm [Fe/H]}}
\newcommand{\logg}{\mbox{${\rm \log g}$}}
\newcommand{\rhk}{\mbox{$\log R^\prime_{\rm HK}$}}
\newcommand{\shk}{\mbox{$S_{\rm HK}$}}
\newcommand{\prot}{\mbox{$P_{\rm rot}$}}
\newcommand{\Mjup}{\mbox{$M_{\rm Jup}$}}
\newcommand{\rchisq}{\mbox{$\chi_{\nu}^2$}}     % reduced chi-square
\newcommand{\Tp}{\mbox{$T_p$}}                  % time of periastron passage
\newcommand{\Msini}{\mbox{$M\sin i$}}           % planet mass
\newcommand{\ms}{\mbox{m s$^{-1}$}}
\newcommand{\msyr}{\mbox{m s$^{-1}$ yr$^{-1}$}}
\newcommand{\kms}{\mbox{km s$^{-1}$}}
\newcommand{\rms}{\mbox{rms}}
\newcommand{\snr}{\mbox{\rm signal-to-noise}}
\newcommand{\caii}{\ion{Ca}{2} H \& K}
  
\received{}
\accepted{}
  
\slugcomment{to appear in ApJ}
  
\shortauthors{Fischer \etal}
\shorttitle{Five Planets}
\begin{document}
  
\title{Five Planets and an Independent Confirmation of HD~196885Ab from  Lick Observatory \altaffilmark{1}}
\author{Debra Fischer\altaffilmark{2,3},
Peter Driscoll\altaffilmark{4},
Howard Isaacson\altaffilmark{2},
Matt Giguere\altaffilmark{2,3},
Geoffrey W. Marcy\altaffilmark{5},
Jeff Valenti\altaffilmark{6},
Jason T. Wright\altaffilmark{7},
Gregory W. Henry\altaffilmark{8},
John Asher Johnson\altaffilmark{9},
Andrew Howard\altaffilmark{5},
Katherine Peek\altaffilmark{5},
Chris McCarthy\altaffilmark{2}
}
  
\email{fischer@stars.sfsu.edu}
  
\altaffiltext{1}{Based on observations obtained at the Lick Observatory,
which is operated by the University of California}
  
\altaffiltext{2}{Department of Physics \& Astronomy, 
San Francisco State University,
San Francisco, CA  94132}

\altaffiltext{3}{Department of Astronomy, 
Yale University, New Haven, CT 06511}
  
\altaffiltext{4}{Department of Earth and Planetary Science, 
Johns Hopkins University, Baltimore, MD}

\altaffiltext{5}{Department of Astronomy, 
University of California, Berkeley, Berkeley, CA}

\altaffiltext{6}{Space Telescope Science Institute, Baltimore, MD}
  
\altaffiltext{7}{Cornell University, Ithaca, NY} 

\altaffiltext{8}{Center of Excellence in Information Systems, 
Tennessee State University,
3500 John A. Merritt Blvd., Box 9501, Nashville, TN 37209}

\altaffiltext{9}{NSF Astronomy \& Astrophysics Postdoctoral Fellow, 
Institute for Astronomy, University of Hawaii,
Honolulu, HI 96822}

\begin{abstract}
We present time series Doppler data from Lick Observatory 
that reveal the presence of long-period planetary companions orbiting 
nearby stars. The typical eccentricity of these massive
planets are greater than the mean eccentricity of known exoplanets. 
HD~30562b has \Msini\ = 1.29 \Mjup, 
with semi-major axis of 2.3 AU and eccentricity 0.76. The host
star has a spectral type F8V and is metal rich. HD~86264b 
has \Msini\ = 7.0 \Mjup, \arel\ = 2.86 AU, an eccentricity,$e = 0.7$ 
and orbits a metal-rich, F7V star. HD~87883b has \Msini\ = 1.78 \Mjup, 
\arel\ = 3.6 AU, $e = 0.53$ and orbits a metal-rich K0V star. 
HD89307b has \Msini\ = 1.78 \Mjup, \arel\ = 3.3 AU, $e = 0.24$ and 
orbits a G0V star with slightly subsolar metallicity.
HD 148427b has \Msini\ = 0.96 \Mjup, \arel\ = 0.93 AU, eccentricity 
of 0.16 and orbits a metal rich K0 subgiant. We also present 
velocities for a planet orbiting the F8V metal-rich binary star, 
HD~196885A. The planet has \Msini\ = 2.58 \Mjup, \arel\ = 2.37 AU, 
and orbital eccentricity of 0.48, in agreement with the independent
discovery by \citet{c08}.

\end{abstract}


\section{Introduction}
  
Over the past 14 years, more than 300 extrasolar planets have been 
discovered orbiting sunlike stars. Most of these discoveries were 
made with high precision Doppler observations that 
measure the reflex radial velocity of the host star. 
Radial velocity amplitudes scale with the mass of the planet
and are inversely proportional to the cube root of the 
orbital period. Therefore, detectability in Doppler surveys is 
enhanced for short period orbits and massive exoplanets. 
An additional constraint on detectability 
is that at least one full orbital period must be observed 
in order to accurately model the Doppler velocity data with a Keplerian 
orbit. As a result, about 85\% of 
detected exoplanets have \Msini\ greater than the mass of Saturn; 
most have orbital periods shorter than four years (e.g., see http://exoplanets.eu)
Higher cadence observations
and improved Doppler precision are now enabling the detection of both lower
mass \citep{bou09, may09a, may09b, how09} and longer period \citep{mou09} exoplanets. 

The planet survey at Lick Observatory began in 1989 and 
is one of the oldest continuous Doppler programs in the world 
\citep{ma97}. Because of this long time baseline of 
data, 55 Cnc d, the only planet with complete 
phase coverage and a well determined orbital period greater 
than ten years was discovered at Lick Observatory \citep{ma02}.  The original 
Lick program contained $\sim 100$ stars. In 1997, the sample was 
augmented to $\sim 400$ stars \citep{fi99}. In this paper, we describe the 
detection of single planets orbiting HD~30562, HD~86264, HD~89307, HD~87883, 
HD~148427, and HD~196885Ab that have emerged from that extended sample. 
All are fairly massive planets (\Msini $> 0.96$ \Mjup) in relatively long 
period (P $> 333$ d) orbits 
that exhibit significant orbital eccentricities.

\section{Doppler Analysis} 
 
Our Doppler analysis makes use of an iodine absorption cell in the 
light path before the entrance slit of the spectrometer.  The iodine 
absorption lines in each program observation are used to model the  
wavelength scale and the instrumental profile of the telescope and
spectrometer optics for each observation \citep{ma92, bu96}
The iodine cell at Lick Observatory has not been 
changed over the entire duration of the planet search project, 
helping to preserve continuity in our velocity measurements despite
three CCD detector upgrades. The velocity precision for the Lick 
project is generally photon-limited with typical \snr\ $\sim 120$. To 
improve the precision, consecutive observations are sometimes independently 
analyzed and a single weighted mean velocity is determined. 

In addition to photon statistics, there are sources of 
systematic errors.  For example, the Hamilton spectrometer \citep{v87} resides in
the Coud\'{e} room and is not temperature controlled. Diurnal temperature 
variations of several degrees can occur in addition to seasonal 
temperature swings of about $\sim$25~C. Temperature and 
pressure changes can lead to gradual changes in the PSF
through the night. Even more rapid PSF variations occur  
as guiding errors or changes in seeing shift the spectrum 
by a few tenths of a pixel (i.e., $\sim$400 \ms) 
on the CCD. Ultimately, the burden for tracking all of 
these systematic sources of error is placed on our ability to model the PSF
from the iodine absorption lines in each program observation. 

Because the PSF varies over the detector, the Doppler analysis is 
carried out independently in wavelength chunks that span 
about 3\AA\ of the spectrum. 
In each of these chunks, we model the wavelength solution, the PSF and the 
Doppler shift of the star. Each chunk is compared differentially 
to chunks from previous observations containing the same spectral lines. 
The median velocity for all the chunks yields the differential 
velocity measurement. The standard deviation of the velocities 
from the several hundred chunks provides an assessment of the uncertainty 
in the velocity measurement, but can be an underestimate of the 
residual RMS scatter arising from instrumental sources or additional 
unidentified planets. 

\subsection{Jitter}
The Doppler analysis provides formal measurement errors, however 
additional systematic noise in the center of mass velocity of the 
star can arise and are more difficult to characterize.  
The systematic errors can either be intrinsic to velocity fields in the 
stellar atmospheres, or, as described in the previous section, 
they can be the result of instrumental effects 
that are inadequately modeled in our Doppler analysis.  Unfortunately, 
it is often difficult to identify the source of spurious velocity variations. 
For example, \citet{q01} measured velocity variations in {HD~165435}
with an amplitude of 80 \ms\ that correlated with photometric variability.
However, systematic studies of activity and velocity variations
\citep{sbm98, sf00} have shown that it is difficult to identify 
unambiguous correlations between astrophysical parameters and 
velocity variations.   

\citet{if09} have measured emission in 
the \caii\ line to produce \shk\ values that are calibrated to 
the canonical Mt Wilson values \citep{d91} following \citet{w05}. 
The \shk\ values
are used to calculate \rhk, the ratio of emmission in the 
cores of the \caii\ lines to the flux in the photosphere, and 
to estimate rotational periods \citep{noyes84}.
They searched for correlations between \rhk\ values 
and radial velocity ``jitter'' using  the  highest precision velocity 
data sets from the California Planet Search survey, obtained after 
the CCD upgrade (August 2004) at Keck Observatory. Their model of 
jitter has a functional dependence on (1) \rhk, (2) height above 
the main sequence (evolutionary status), and (3) \bv\ color. 
We adopt their model as an independent estimate of jitter (which 
likely includes some systematic instrumental errors in addition to astrophysical 
noise) that is added in quadrature to formal velocity errors when fitting 
Keplerian models to the data.  These augmented errors are included 
in the calculation of \rchisq\ and in the Figures showing Keplerian 
fits to the data; however, they are not included in the tabulated 
velocity errors for the time series data. To reproduce the 
errors used in Keplerian fitting, the stellar jitter values 
should be added in quadrature to errors reported in the individual
radial velocity Tables.

\section{Keplerian Fitting} 

\subsection{Levenberg Marquardt Fitting}  
Time series velocity data are fit with a Keplerian model 
using the partially linearized Levenberg-Marquardt minimization algorithm
described in \citet{wh09}.  
The Levenberg-Marquardt algorithm employs a gradient search to 
minimize the \rchisq\ fit between an assumed Keplerian model 
and the observed data. The free parameters for each planet in the 
Keplerian model include the orbital period, $P$, the time of 
periastron passage, $T_P$, the orbital eccentricity, $e$, the 
argument of periastron passage referenced to the ascending 
node, $\omega$ and the velocity semi-amplitude, $K$.  The 
velocity for the center of mass, $\gamma$, is an additional single 
free parameter for a system of one or more planets. The Keplerian 
model with the minimum \rchisq\ value provides the maximum 
likelihood estimate for the orbital parameters. While this 
approach is quite efficient for well-sampled data sets,
it is possible for the fitting algorithm to become trapped 
in a local \rchisq\ minimum. Particularly when there are unobserved gaps 
in the orbital phase, covariance between the orbital 
parameters can be significant and Keplerian fitting with a 
Levenberg-Marquardt algorithm does not capture the full 
range of possible orbital parameters. 

Typically, $N=1000$ Monte Carlo trials are run with Levenberg-Marquardt
Monte Carlo (LMMC) fitting, scrambling the residuals 
(with replacement of previously selected values) 
before adding the theoretical velocities back in and 
refitting new Keplerian models using the Levenberg-Marquardt
algorithm. For Gaussian velocity errors, mean values 
from the Monte Carlo trials define 
the parameter value. The standard deviation of the trial 
parameters defines the uncertainties. 

\subsection{Markov Chain Monte Carlo Fitting}
Following a technique outlined by \citet{f05} we have also tested  
a Markov Chain Monte Carlo (MCMC) algorithm to derive Keplerian 
fits and to characterize uncertainties in the orbital parameters
\citep{dff09}. The MCMC algorithm is a 
Bayesian technique which samples the orbital parameters in 
proportion to an expected posterior probability distribution. It allows 
for larger steps in parameter space than a routine driven 
only by \rchisq\ minimization, enabling a more complete 
exploration of parameter space.

One drawback of the MCMC method is that, since each step in the
Markov chain is correlated with the previous step, MCMC can give
misleading results if the Markov chain has not converged.  
For systems that do converge, the MCMC method provides a more 
robust characterization of the uncertainties in orbital parameters, 
particularly when the observational data results in either a 
rough \rchisq\ surface or a \rchisq\ surface with a shallow 
minimum. For data sets where the observations 
span multiple orbital periods with complete phase coverage, 
the MCMC algorithms converge to the same solutions
obtained by the frequentist Levenberg-Marquardt algorithm 
\citep{dff09}.

\section{HD 30562}
  
\subsection{Stellar Characteristics}
  
HD~30562 (HIP 22336, $V=5.77$, \bv = 0.63) is a F8V star.
The Hipparcos parallax \citep{esa97} yields a distance of 
26.5 pc. \citet{if09} find that HD~30562 is chromospherically 
inactive and measure \shk\ $=0.15$, \rhk\ = -5.064. Based on their
model of stellar jitter, we estimate a jitter of 2.9 \ms\ for 
this star. We estimate a rotational period, \prot\ = 24.2 d, using the 
calibration by \citet{noyes84}. 

A high resolution spectroscopic analysis has been carried out 
for all of the stars in this paper, including HD~30562, using 
spectroscopic modeling with SME \citep{vp96}.  The 
analysis that was reported in \citet{vf05} has been 
further refined, following the method described in \citet{v09}:
the SME spectroscopic solution for surface gravity is determined 
iteratively with interpolation of the Yonsei-Yale (``Y2'') 
isochrones \citep{d04}. 
This analysis yields 
$\teff= 5861 \pm 44$K, 
$\logg=4.088 \pm 0.06$, $\vsini\ =4.9\pm0.50$ \kms, 
$\fe=0.243\pm0.04$ dex. 
The isochrone analysis yields the same value for surface gravity (by design) 
and a stellar luminosity of $\lstar = 2.85\lsun$ with a 
bolometric correction of $\bc=-0.064$, a stellar mass of 
$1.219 \msun$ and a stellar radius of 1.637 \rsun.
This compares well with the stellar mass and luminosity 
derived by \citet{t07} who model stellar evolution 
tracks using the spectroscopic model parameters. 
The stellar parameters described here are summarized 
in Table \ref{tab_t1}. 

\subsection{Doppler Observations and Keplerian Fit}
  
We have acquired 45 Doppler measurements of 
HD~30562 over the past ten years. 
With typical seeing at Lick of $1\farcs 5$, the 
exposure time for SNR of 140 is about 5 minutes on the 
3-m Shane telescope or about 30 minutes on the 0.6-m Coud\'{e}
Auxiliary Telescope (CAT). 

The observation dates, radial velocities and measurement 
uncertainties are listed in Table \ref{tab_t2}.  
The initial phase coverage was rather poor, however, 
after velocity variations were detected we redoubled our 
efforts and obtained additional (co-added) observations using 
the 0.6-m CAT to fill in 
phase coverage since 2006. The best fit LMMC Keplerian model 
has a period of $P = 1157 \pm 27$ days, a semi-velocity 
amplitude $K = 33.7 \pm 2.2$ \ms\ and orbital eccentricity, 
$e = 0.76\pm 0.05$. 
The mean RMS to the fit is 7.58 \ms. 
Including the estimated jitter of 2.9 \ms\ 
we obtain \rchisq\ = 1.31 as a measure of the goodness 
of the model fit. Adopting a stellar mass of 1.219 \msun, 
we derive \Msini\ = 1.29 \Mjup\ and a semi-major axis of 2.3 AU. 

\begin{figure}
\plotone{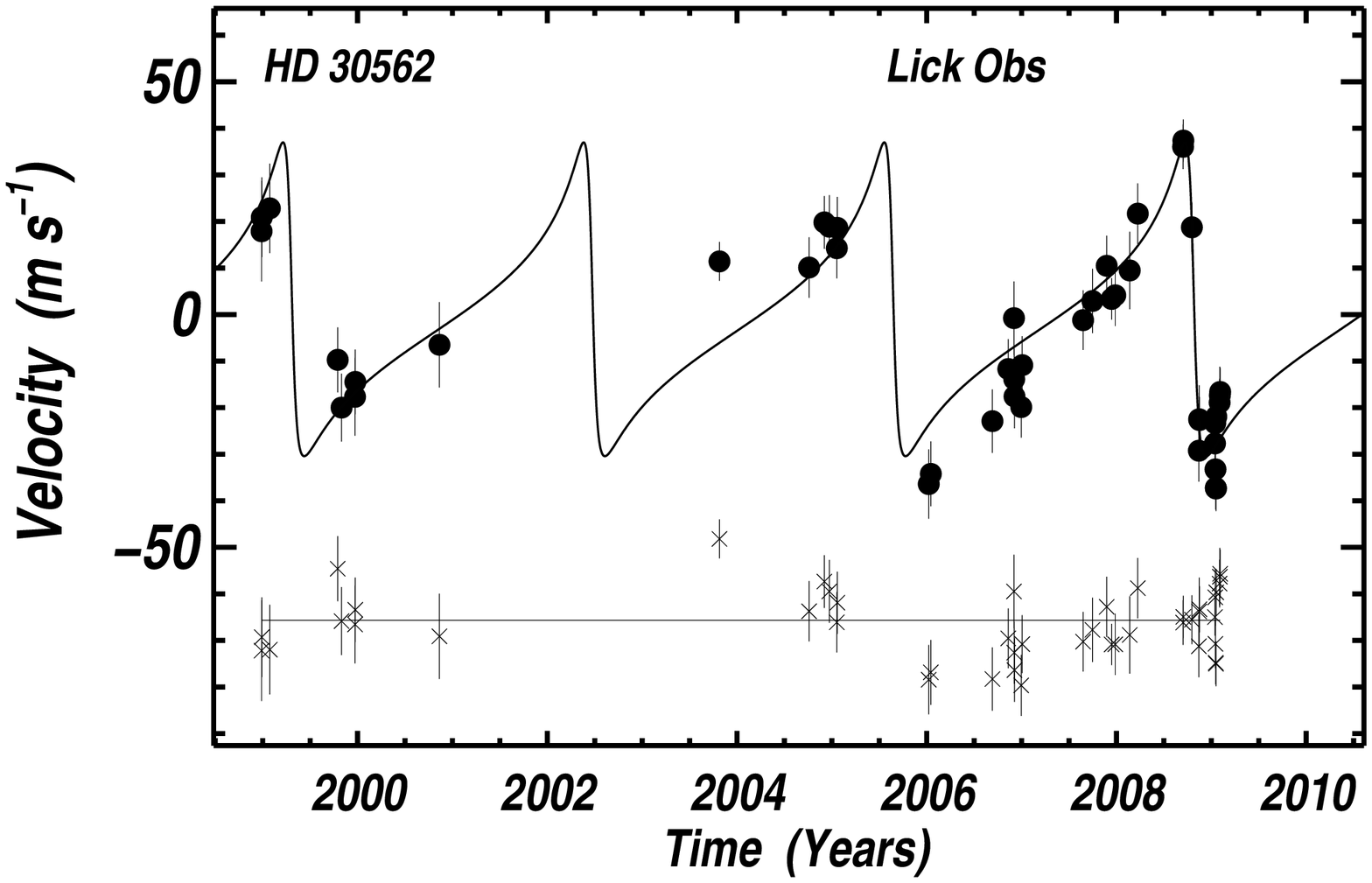}
\figcaption{Time series radial velocities from Lick 
Observatory are plotted for HD 30562 with
2.9 \ms of expected velocity jitter added in 
quadrature with the single measurement uncertainties. 
The Keplerian model is overplotted with 
an orbital period of 3.2 years, velocity amplitude 
of 33.7 \ms and eccentricity, $e = 0.76$. 
With these parameters and the stellar mass of 1.219 \msun, 
we derive a planet mass, \Msini\ = 1.29 \Mjup\ and semi-major
axis of 2.3 AU.  Residual velocities to the fit are offset and 
show some slight systematic variation.  
\label{fig_rvfit_30562}}
\end{figure}

As described above, uncertainties in the orbital parameters were 
determined by running 1000 LMMC trials.  In each trial, the theoretical 
fit was subtracted from radial velocities and the residual velocities 
were scrambled and added back to the theoretical velocities.
A new trial Keplerian fit was then obtained. The standard deviation 
of each orbital parameter for the 1000 Monte Carlo 
trials was adopted as the parameter uncertainty. 
The Keplerian orbital solution is listed in 
Table \ref{tab_t3} and the time 
series velocity data are plotted with the best-fit Keplerian 
model (solid line) in Figure \ref{fig_rvfit_30562}.

\begin{figure}
\plotone{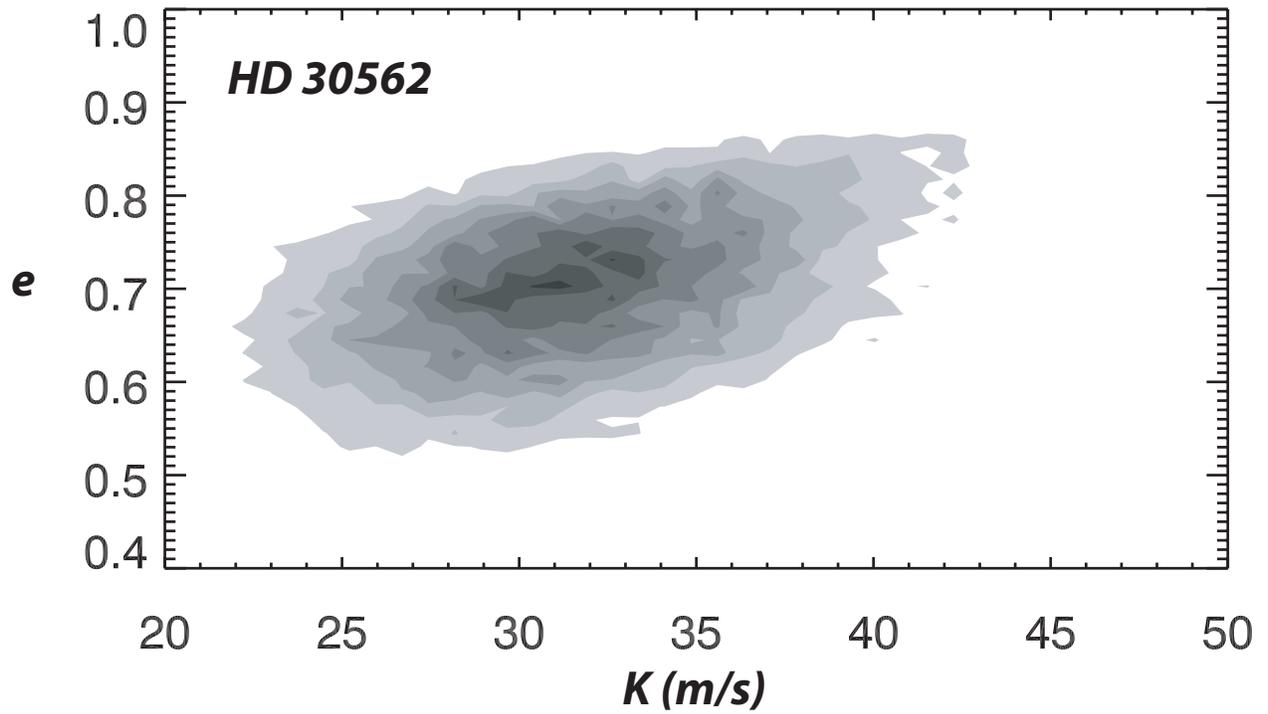}
\figcaption{The MCMC analysis revelas a modest covariance between 
the orbital eccentricity and velocity amplitude that  
is a factor of two larger than the formal errors from the LMMC analysis.  
\label{fig_cov_30562}}
\end{figure}

We also carried out an MCMC fit for HD 30562. The radial velocity
data have good phase coverage for this system, and so the 
agreement between the LMMC and MCMC fits are quite good.
The modest covariance between the velocity amplitude and 
orbital eccentricity is shown in Figure \ref{fig_cov_30562},
however, the best \rchisq\ contours are consistent 
with the formal parameter errors derived from the LMMC analysis.

\section{86264} 

\subsection{Stellar Characteristics}

HD 86264 (HIP 48780) is an F7V star with apparent brightness, 
$V = 7.42$, and color \bv\ = 0.46. Based on the parallax 
measurement from Hipparcos \citep{esa97} of $13.78 \,mas$, 
this star is located at a distance of about 72 parsec with an 
absolute visual magnitude of \mv\ = 3.10. Using spectral synthesis
modeling and iterating until there is a match in surface gravity
with the value predicted from interpolation of the Y2 isochrones,
we derive $\teff = 6210 \pm 44$K, \logg\ = 4.02, \vsini\ = 12.8 \kms,
\fe\ = +0.202.  The bolometric correction is -0.024, stellar luminosity
is \lstar\ = 4.55 \lsun, \rstar\ = 1.88 \rsun, and \mstar\ = 1.42 \msun.
The star is moderately active, with \shk\ = 0.20 and \rhk\ = -4.73. 
The expected jitter from \citet{if09} is 3.3 \ms.
The stellar characteristics are compiled in Table \ref{tab_t1}.

\subsection{Doppler Observations and Keplerian Fit}

HD 86264 has been on the Lick program since
January 2001 and exhibits a large amplitude velocity variation 
with a periodicity of about 4 years. The expected jitter 
for this star is likely higher than predicted 
because the star is a relatively rapid rotator.
For this reason, and because the 
star is at the faint magnitude limit of our program, we 
typically limited our exposure time and obtained 
SNR of only 80 to 100. Our mean velocity precision 
for HD 86264 is 21 \ms\ (probably set 
both by the relatively low SNR for our observations and the 
relatively high vsini of the star). As a result, only 
planets with relatively large velocity amplitudes would have been detected 
around this star. 

A total of 37 radial velocity measurements are listed 
in Table \ref{tab_t4}. A periodogram of the velocities shows a 
strong broad peak at about 1475 days with an FAP
less than 0.0001 or 0.01\%. The Keplerian model was derived 
with a Levenberg-Marquardt (LMMC) fitting algorithm with an assumed 
stellar jitter of 3.3 \ms\ added in quadrature to the formal 
velocity errors. 
The best fit LMMC solution has a period of $P = 1475 \pm 55$d,
velocity amplitude, $K=132$ \ms, and eccentricity $e = 0.7$. 
However, the LMMC trials revealed an asymmetry in the distribution 
of modeled velocity amplitudes. While $K$ was rarely less than 120 \ms,
some models were found with $K$ up to 246 \ms. Furthermore, 
a large standard deviation in the 1000 trials was found for 
the orbital eccentricity. 
The \rchisq\ fits for eccentricities down to 0.4 were only 
worse by 5\%, with \rchisq\ = 1.28. 

The linear trend included in the LMMC Keplerian model has a 
positive slope of about 1.8 \ms\ per year
or 16.4 \ms\ over the nine year duration 
of velocity measurements. This slope is only marginally significant given the 
large uncertainty in the radial velocity measurements for this star, however 
it was retained because of the significant improvement to \rchisq. 
We note that a similar improvement in \rchisq\ could have been achieved
by adopting a larger (and still physically plausible) value for jitter.

\begin{figure}
\plotone{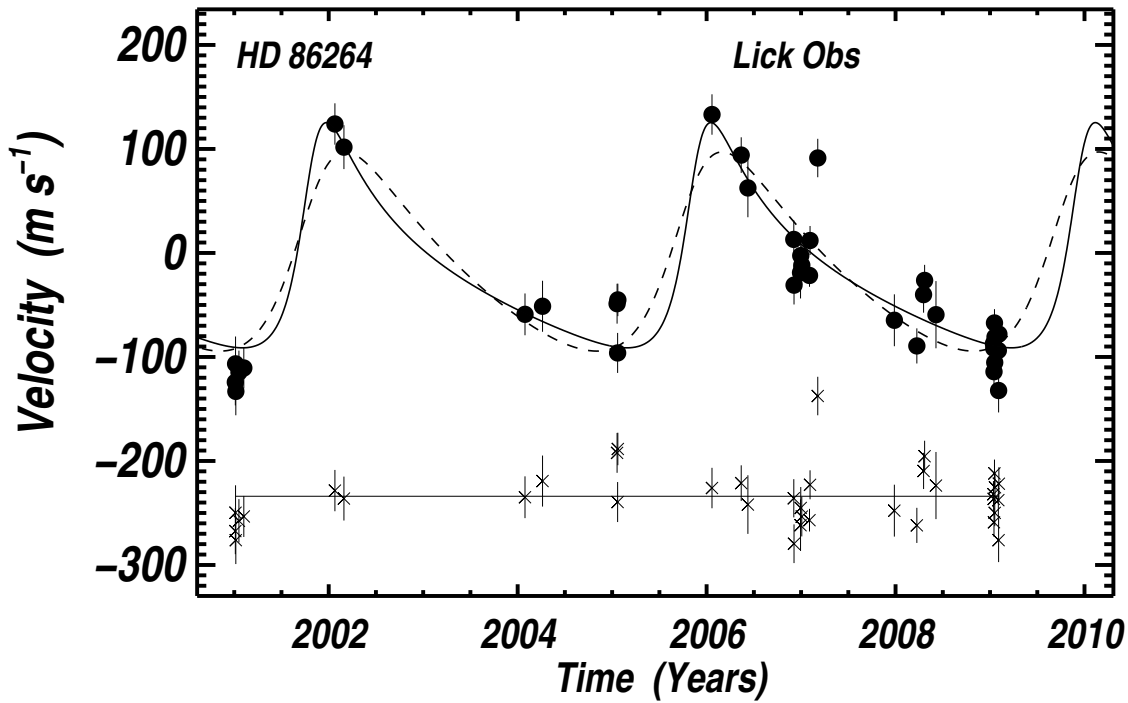}
\figcaption{Time series radial velocities for HD 86264 from 
Lick Observatory.
The best fit Keplerian orbit is plotted as a solid line 
and has an orbital period of 
almost 4 years, a velocity amplitude of 132 \ms and eccentricity 
of 0.7. The stellar mass of 1.42 \msun yields 
a planet mass, \Msini\ = 7 \Mjup\ 
and semi-major axis of 2.86 AU. An alternative Keplerian 
model with an eccentricity of 0.4 is shown with a dashed line. 
Such a model has a \rchisq\ fit that is only worse by 5\% 
compared to the 0.7 eccentricity solution.  
\label{fig_rvfit_86264}}
\end{figure}

The implied planet mass is \Msini\ = 7 \Mjup\ with 
a semi-major axis of 2.86 AU. The LMMC Keplerian solution 
is summarized in Table \ref{tab_t3}. 
The time series velocity measurements are plotted in 
Figure \ref{fig_rvfit_86264}. The LMMC Keplerian model with 
best fit eccentricity of 0.7 is indicated in 
Figure \ref{fig_rvfit_86264} with a solid 
line and a Keplerian model with eccentricity of 0.4 is 
overplotted with a dashed line.  Although the difference 
between 0.4 and 0.7 is substantial,
it is apparent from Figure \ref{fig_rvfit_86264} 
that the solutions are nearly consistent with either  
value, resulting in a modest 5\% penalty in \rchisq.

\begin{figure}
\plotone{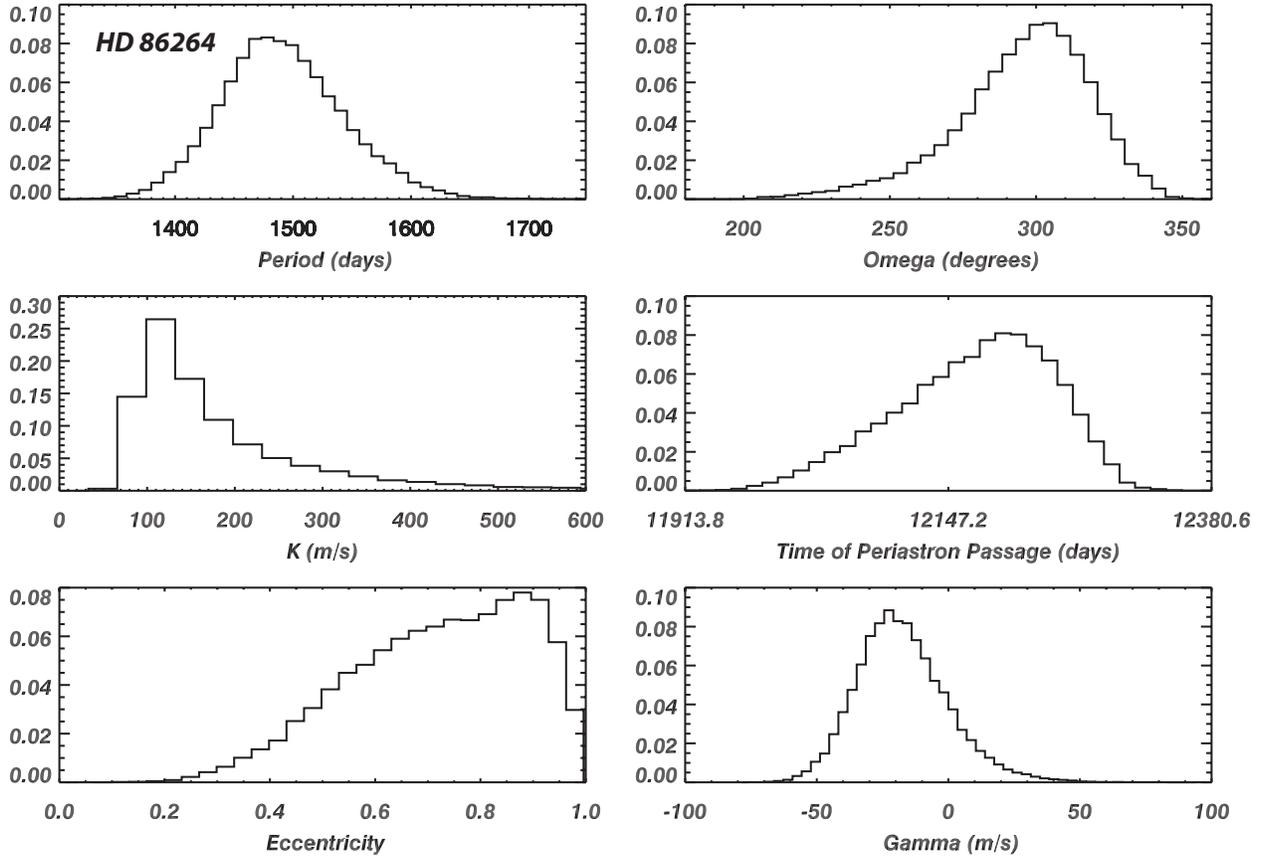}
\figcaption{The posterior probability distributions from 
Markov Chain Monte Carlo (MCMC) simulations show peak values 
close to those derived from Levenberg-Marquardt fitting.
The widths of these distributions characterize the uncertainty in 
the orbital parameters.
\label{fig_pdf_86264}}
\end{figure}

\begin{figure}
\plotone{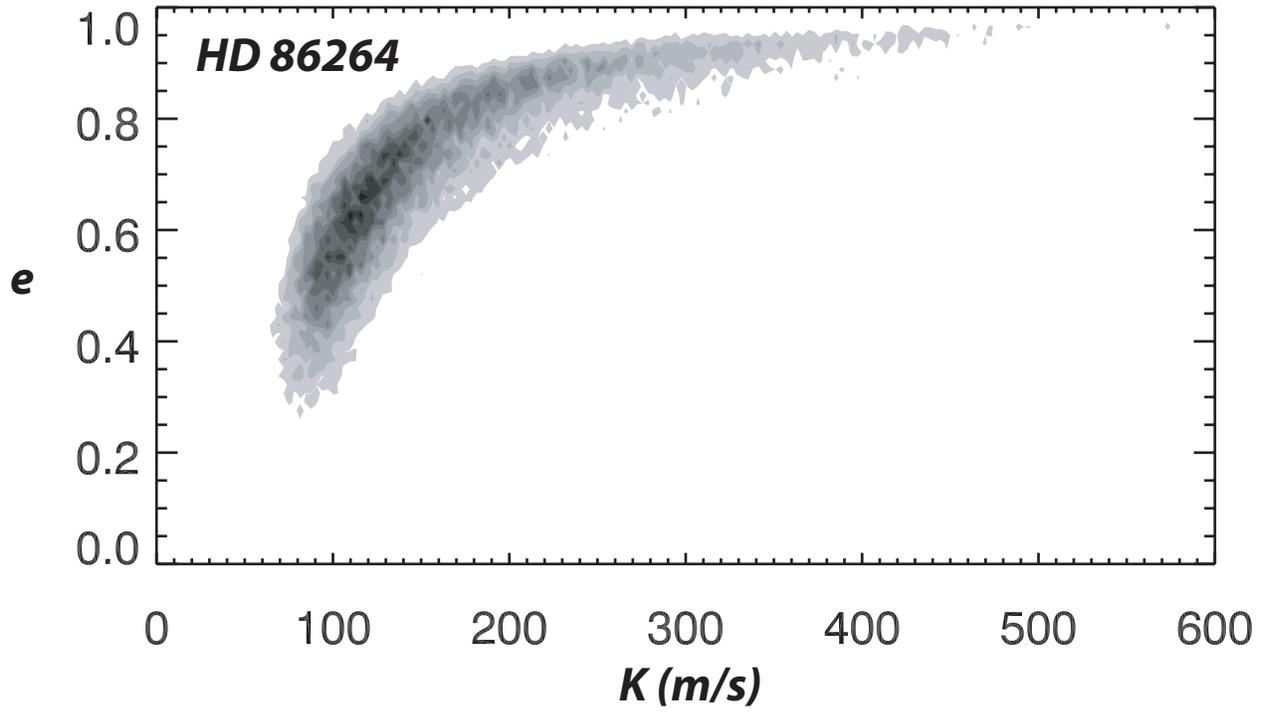}
\figcaption{The MCMC probability density functions
used to model the Keplerian orbital parameters  
reveals covariance between orbital 
eccentricity and velocity amplitude. For the most 
likely velocity amplitude of about 132 \ms, eccentricities 
between 0.5 and 0.7 are plausible. 
\label{fig_cov_86264}}
\end{figure}

The radial velocity data set for HD 86264 has a gap 
in the time series data as the planet approaches periastron. 
As a result, the LMMC fitting algorithm may not have 
captured the full range of possible parameters (also suggested 
by the large uncertainty in eccentricity and velocity amplitude). 
The probability density functions from the MCMC simulations are 
plotted in Figure \ref{fig_pdf_86264} and show general agreement 
with the LMMC trials.  However, the MCMC simulations quantify a 
broader range of parameter values, particularly for  
orbital eccentricity and the velocity amplitude. The covariance of  
these two parameters is illustrated in Figure \ref{fig_cov_86264}; 
eccentricity is correlated with velocity amplitude because of the 
gap in the time-series velocity measurements. 

\subsection{Photometry}
HD~86264 is the only star in the present sample for which we have 
photometric observations.  The star was observed in the Johnson $B$ and $V$
pass bands by the T3 0.4~m automatic photometric telescope (APT) at Fairborn 
Observatory as one of two comparison stars for another observing program 
\citep[][Tables 3 and 4]{hfh2007}.  Details of the T3 APT, the differential 
observing sequence, and the reduction of the data are given in that same 
paper.

Between 2003 November and 2004 May, the T3 APT acquired 239 good observations
in the $V$ band and 232 observations in $B$.  We have reanalyzed these
observations from \cite{hfh2007} for the present study, searching for 
low-amplitude variability that might allow the direct determination of the 
star's rotation period.  With a log R'HK value of -4.73, HD~86264 is the 
most active star in our sample and so a good candidate for exhibiting starspot 
activity, which might result in detectable rotational modulation of the 
star's brightness \citep[][see Figure~11]{h1999}. 

The {\it Hipparcos} catalog \citep{esa97} lists 107 photometric
measurements acquired during the mission between 1989 November and 1993 
March, but the {\it Hipparcos} team does not venture a variability 
classification.  Our 2003--04 APT measurements have standard deviations of
0.0065 and 0.0064 in the $B$ and $V$, respectively.  These values agree
with typical scatter observed in constant stars measured with the T3 APT.
Power spectrum analysis fails to find any significant periodicity between 
one and 25 days and limits the semi-amplitude of any real signal within
this period range to a maximum of $\sim0.0015$ mag.

We were somewhat surprised by our failure to detect rotational variability
in HD~86264, given its modest activity level.  However, we note from 
Tables 1 and 8 that HD~86264 has the lowest color index in the sample 
($B-V=0.46$) and thus a relatively shallow convection zone.  Furthermore,
the {\it estimated} rotation period, stellar radius, and observed 
$v$~sin~$i$ given in Table~1 imply a low equatorial inclination of 
$\sim30^\circ$.  Both of these factors work to minimize observable rotational
variability in HD~86264.  Surface magnetic activity should not have a 
significant effect on the measured radial velocities.

\section{87883}

\subsection{Stellar Characteristics}

HD 87883 (HIP 49699) is a K0V star with $\bv\ = 0.96$ 
and Hipparcos parallax based distance of 18 parsecs. The star 
has a V magnitude of 7.57 and an absolute visual magnitude, 
\mv\ = 6.3. \citet{if09} measure modest chromospheric activity with 
an \shk\ value of 0.26, \rhk\ = -4.86 and predict a stellar jitter 
of 4.5 \ms. The rotation period predicted from this activity 
level is 41.2 days \citep{noyes84}.
We again adopt an iterative approach to tie surface gravity 
from the spectroscopic analysis to the Y2 isochrone interpolation. 
The surface gravity converges at \logg\ = 4.58 and yields 
$\teff\ = 4980 \pm 44$K, \fe\ = +0.093, \vsini\ = 2.2 \kms 
in good agreement with the original analysis of \citet{vf05}.
The stellar luminosity is \lstar\ = 0.318 \lsun, stellar radius is 
\rstar\ = 0.76 \rsun, and stellar mass is 0.82 \msun.
The stellar parameters are summarized in Table \ref{tab_t1}. 

\subsection{Doppler Observations and Keplerian Fit}

HD 87883 has been observed at Lick Observatory since 
December 1998. The 44 radial velocity measurements from Lick 
Observatory are listed in 
Table \ref{tab_t5}, along with the observation dates 
and uncertainties. The mean \snr\ of 120 for the observations 
produces a typical velocity measurement uncertainty of 4 \ms. 
The data were initially fit with a single planet model with a 
period of 7.9 years. The initial \rms\ of the Keplerian 
fit was surprisingly high, 9.0 \ms. After adding 
the expected jitter of 4.5 \ms\ in quadrature with the 
internal errors, we found a relatively poor \rchisq\ of 1.7, 
suggesting that the single planet model was not adequate. 
The velocity residuals to the prospective Keplerian model 
of the Lick data showed only a modest 
peak in the periodogram. 

In an effort to better understand the residual velocities, 
we obtained 25 additional velocity measurements from the 
Keck Observatory. These velocities are included in 
Table \ref{tab_t5} with 
a designation of ``K'' in the ``Observatory'' column to 
distinguish them from the Lick observations. The average \snr\ 
at Keck is 220 and the single measurement precision is $\sim 2.2$ \ms,
providing higher quality velocity measurements. 
The Lick and Keck velocities were merged with velocity offset as
a free parameter to minimize \rchisq\ of the Keplerian fit. 
The velocity offset was only 1.6 \ms\ for these two data sets 
and has been subtracted from the Keck radial velocity measurements 
listed in Table \ref{tab_t5}.

\begin{figure}
\plotone{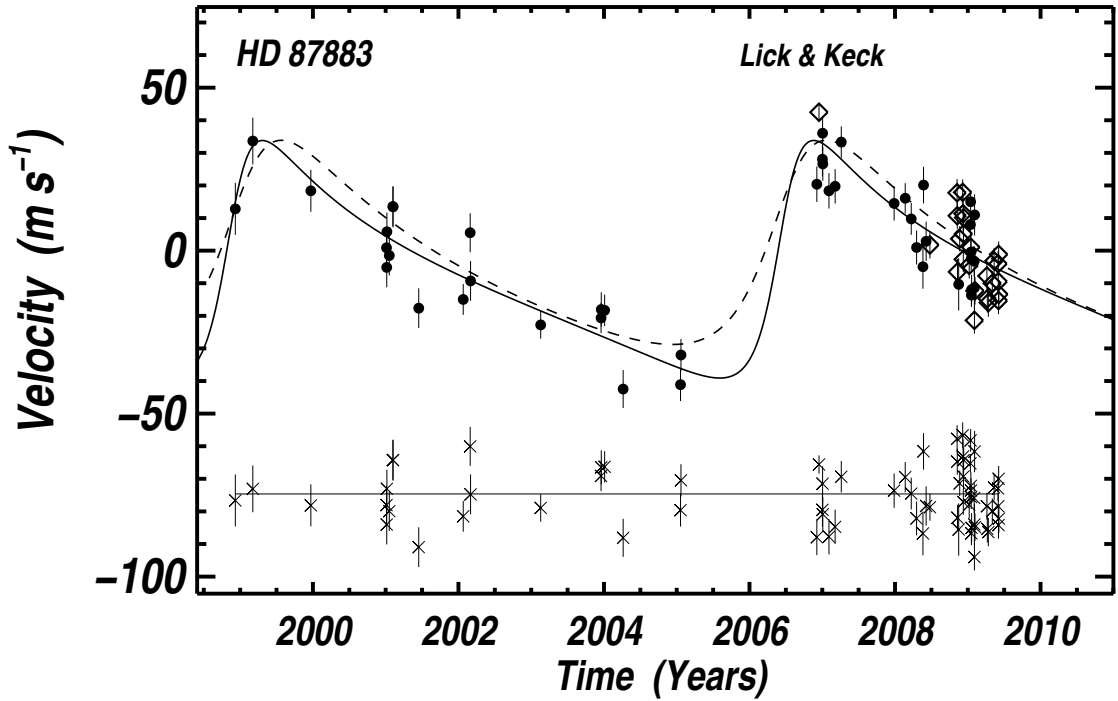}
\figcaption{Our time series Doppler measurements 
from Lick Observatory are shown with filled circles 
and velocities from the Keck Observatory are 
represented by diamonds. We have added 4.5 \ms velocity 
jitter in quadrature to both of the error bars shown 
here. The best fit Keplerian orbital period is 7.6 years, the 
eccentricity for this planet is $e = 0.53$ 
and the velocity semi-amplitude is $K=34.7$ \ms.
Because there is a gap in the phase approaching periastron, 
the eccentricity is poorly constrained and values as low 
as 0.4 are plausible for this star. 
With the stellar mass of 0.82 \msun\ we derive a planet mass, 
\Msini = 1.78 \Mjup\ and semi-major axis of 3.6 AU.
The best fit theoretical curve is overplotted with a
solid line and the lower eccentricity solution is plotted with 
a dashed line. 
\label{fig_rvfit_87883}}
\end{figure}

The combined Lick and Keck velocities were modeled with a 
Keplerian with a period of 7.9 years, an eccentricity of $e = 0.53$, and 
velocity semi-amplitude of $K = 34.7$ \ms. The velocities are 
plotted in Figure \ref{fig_rvfit_87883} with the expected 4.5 \ms\ jitter 
added in quadrature with the formal errors. 
After fitting the combined Lick and Keck velocities, the \rms\ for 
the Keck data alone was 8.6 \ms\ and the \rms\ for 
the combined data sets was 9.2 \ms\ with a \rchisq\ = 1.71. 
The periodogram of the residuals to the fit (with combined Lick and 
Keck data) does not show any significant power. Continued observations
may eventually reveal an additional short-period planetary companion. 
Alternatively, a background star, blended with the 
image of HD~87883 on the slit could also introduce unexpected 
velocity variations. Since the star is relatively close (18 parsecs) 
a stellar companion separated by less than 0\farcs 5 might be 
resolvable with adaptive optics observations and would be helpful 
for understanding the high \rms\ to our Keplerian fit. 

\section{89307}
  
\subsection{Stellar Characteristics}
  
HD 89307 (HIP 50473) is a G0V star with an apparent magnitude of 
V = 7.06 and $\bv\ = 0.640$. The Hipparcos-based distance 
is 30.9 parsecs implying an absolute visual magnitude of \mv\ = 4.57. 
Our spectroscopic analysis yields
\teff\ = 5950 $\pm$ 44K, 
\logg\ = 4.414 $\pm$ 0.10,
\vsini\ = 3.21 $\pm$ 0.50 \kms, \fe\ = -0.14 $\pm$ 0.04 dex.
The \logg\ value in the spectroscopic model was tied to the Y2
isochrones, which yield 
a stellar luminosity of 1.24 \lsun\ with a 
bolometric correction of -0.075, radius of 1.05 \rsun\ 
and stellar mass of 1.028 \msun, in good agreement with 
\citet{t07} who derive a stellar mass
of 0.989 \msun, age of 6.76 Gyr, a stellar radius of 1.1 
\rsun\ and \logg\ of 4.36.

{HD 89307} is chromospherically inactive with a measured $\shk\ = 0.154$, 
\rhk\ = -4.98, and estimated velocity jitter of 2.8 \ms.
The inferred rotational period from the chromospheric activity is 23.7d. 
The stellar properties of HD 89307 are compiled in Table \ref{tab_t1}. 

\subsection{Doppler Observations and Keplerian Fit}

We obtained 59 observations of HD 89307 with a typical 
SNR of 120 using the Shane 3m telescope at Lick Observatory 
over the past ten years, yielding a mean velocity precision of 
about 6 \ms. The orbital solution was presented before the 
orbital solution was secure in \citet{fv05} with an orbital 
period of 3090 days and in \citet{bu06} with an orbital 
period of $2900 \pm 1100$ d
ays.

The observation dates, radial velocities and instrumental 
uncertainties for this system are listed in Table \ref{tab_t6}. 
The time series data are plotted in Figure \ref{fig_rvfit_89307} and exceed 
more than one full orbit. The data were fit with 
a Keplerian model using a Levenberg-Marquardt algorithm (LMMC).  
The best fit orbital solution has a period, $P = 2157 \pm 63$ d; 
eccentricity, $e = 0.241 \pm 0.07$; and velocity semi-amplitude, 
$K = 28.9 \pm 2.2$ \ms. With the assumed stellar mass 
of 1.028 \msun\ for this slightly metal-poor star, 
we derive a planet mass, \Msini\ = 1.78 \Mjup\ and semi-major 
axis of 3.27 AU. The \rchisq\ for this fit is 1.37 with an 
RMS of 9.9 \ms. Orbital parameters for HD 89307 are listed 
in Table \ref{tab_t3}.

\section{HD 148427}

\subsection{Stellar Characteristics}

HD 148427 (HIP 80687) is a moderately evolved K0 subgiant 
with an apparent brightness $V = 6.9$ and $B - V$ color $0.98$.  
A distance of 59.3 parsec was derived from the Hipparcos 
parallax of 16.87 $mas$, which yields an absolute magnitude 
\mv\ = 3.02 and luminosity of 6 \lsun\ for this star.  
Spectroscopic modeling of the star provides 
\teff\ = 5052 $\pm$ 44K, \logg\ = 3.586 $\pm$ 0.10, 
\vsini\ = 2.13 \kms\ and \fe\ = 0.154 $\pm$ 0.04. 
Our iterative interpolation of the Y2 isochrones yields
a stellar mass of 1.45 \msun, stellar radius 
of 3.22 \rsun in good agreement with \citet{t07} who 
also derive an age of 2.5 Gyr from evolutionary tracks. 
\citet{if09} derive \shk\ = 0.14, \rhk\ = -5.18 and a stellar 
jitter of 3.5 \ms. 
The stellar parameters are summarized in Table \ref{tab_t1}.

\subsection{Doppler Observations and Keplerian Fit}

HD 148427 has been observed at Lick since 2001 with a 
typical SNR of 120 and single measurement 
uncertainties of about 4 \ms. The radial velocity 
observations of this star are listed in Table \ref{tab_t7}. 
The periodogram of these velocities has a strong peak 
at about 331 days with an FAP less than 0.01\%.  
The data are well-modeled with a Keplerian orbit that 
has a period of $331.5 \pm 3.0$ days, velocity semi-amplitude of 
27.7$\pm 2$ \ms, and 
an eccentricity of $0.16 \pm 0.08$. 
Adopting the stellar mass of 1.45 \msun, we derive a planet mass, 
$\Msini\ = 0.93$ \Mjup\ and an earthlike 
orbital radius of 0.96 AU. The orbital solution is 
listed in Table \ref{tab_t3}. The radial velocity measurements 
are plotted with jitter of 3.5 \ms\ added in quadrature in 
Figure \ref{fig_rvfit_148427} to yield a \rchisq\ fit of 
1.08 with an \rms\ of 7.0 \ms. 
The phase-folded Keplerian model is overplotted in Figure \ref{fig_rvfit_148427}
as a dashed line.

\section{HD 196885 A}
  
We began observing {HD 196885 A} at Lick Observatory in 1998. 
In 2004 the velocity variations for this star were modeled with 
a preliminary orbital period of $P = 346$d.  This (unpublished) result 
appeared temporarily on the California Planet Search 
exoplanet website (as noted by 
\citet{c08}) and was picked up on the Extrasolar Planets 
Encyclopedia \citep{sch09}. However, when it became apparent 
that a significant residual trend had skewed the Keplerian 
model, the link was removed from 
our website while additional data were collected. Although 
the initial fit was incorrect, one advantage of this early 
notice to the community is that the star was 
added to the NACO direct imaging survey at the VLT \citep{ch06} 
and a low mass stellar companion was imaged, 
HD 196885 B, with an angular separation of only 0\farcs7 
corresponding to a projected linear separation of 23 AU. \citet{ch07}
report that photometry of the stellar companion is consistent 
with an M1V dwarf star with a mass of 0.5 - 0.6 \msun. 

\subsection{Stellar Characteristics}

{HD 196885 A} (HIP101966) is a F8V star with absolute visual 
magnitude \mv\ of 3.79.
The apparent stellar magnitude is V=6.39, and color is \bv\ = 0.509. 
The Hipparcos parallax is 0.0303 arcseconds, placing this star 
at a distance of 33 parsecs. We obtained a spectroscopic solution, 
iterating to obtain the same value for \logg\ in the Y2 isochrones.
We measure \teff\ = 6254 $\pm$ 44K, \logg\ = 4.31 $\pm$ 0.1,
\vsini\ = 7.8 $\pm$ 0.50 \kms, and \fe\ = 0.22 $\pm$ 0.04 dex. 
Including a bolometric luminosity correction of -0.028, we obtain 
a stellar luminosity of 2.4 \lsun\
from the Y2 stellar evolutionary tracks, with a 
radius of 1.31 \rsun\ and mass of 1.28 \msun.
The stellar age from evolutionary tracks 
is 3.12 (2.72, 3.48) Gyr \citep{t07}. \citet{if09} measure 
\shk\ = 0.148 and \rhk\ = -4.98, indicating low chromospheric activity for 
{HD 196885 A}. Based on the activity level, we estimate  
a stellar rotation period of 9.4 days and intrinsic radial velocity 
jitter of 2 \ms. Stellar parameters for {HD 196885 A} 
are summarized in Table \ref{tab_t1}.

\subsection{Doppler Observations and Keplerian Fit}

As noted at the beginning of this section, HD 196885 A has an M dwarf
stellar companion with a projected separation of only 23 AU. 
The time series velocities are plotted in 
Figure \ref{fig_rvfit_196885} and show an obvious large amplitude 
trend with curvature, caused by the stellar companion. The reflex 
velocities from the stellar orbit are modulated 
by a lower amplitude variation from a planet orbiting the 
primary component of this binary star system.  
 
The observation dates, radial velocities, and uncertainties for 
{HD 196885 A} are listed in Table \ref{tab_t8}. Seventy-five 
observations have been obtained at Lick Observatory since 1998 July. 
Figure \ref{fig_rvfit_196885} shows the time series radial velocity 
measurements with a Keplerian model that is the combination of a planet 
model plus a stellar binary orbit. In fitting the Keplerian 
orbit, we added 2 \ms\ in quadrature to the internal errors as the best 
estimate for stellar noise based on the spectral type and 
activity of the star. However, we note that the close M dwarf
companion will contaminate the spectrum of the primary star, 
increasing our Doppler errors. 
We first tested periods from 30 to 100 years for the stellar 
companion detected by \citet{ch07}.
With a double-Keplerian model for the binary star and planet, 
we found that \rchisq\ decreases to a minimum of 1.43 for orbital 
periods greater than about 40 years. However, \rchisq\ is flat for 
longer orbital periods, out to $\sim$200 years with strong covariances 
in the solutions in the orbital elements ($K$ and the period) of 
the stellar binary system. 

The best fit Keplerian solution for the planetary orbit, {HD 196885 Ab}, 
was determined with a Levenberg-Marquardt search of parameter space, which is reliable
with phase coverage spanning several orbits.  Our 
best fit model has an 
orbital period of $1333 \pm 15$d, eccentricity $0.48 \pm 0.06$, 
velocity semi-amplitude $K = 53.9 \pm 3.7$. The residuals to the 
fit of {HD 196885 Ab} have an RMS of 14.7 \ms\ and \rchisq\ = 1.58.
The Keplerian solution for the planet orbit is summarized in 
Table \ref{tab_t3}.

Figure \ref{fig_rvfit_196885c} shows the modeled Keplerian 
orbit for {HD 196885 Ab} 
after the orbit from the assumed stellar binary {HD 196885 B}
companion has been subtracted off. The contribution of light 
from the spatially unresolved M dwarf companion should be less than 
1 part in 1000, but may have added systematic errors in our 
Doppler analysis that resulted in the poorer fit.  
Figure \ref{fig_rvfit_196885c} shows the (stellar binary) residual 
velocities after the planetary orbit from {HD 196885 Ab} is removed.
Clearly, the fractional phase of the observed stellar is not enough 
to constrain the period, amplitude or eccentricity of its orbit. 

{HD 196885} was also observed with ELODIE and CORALIE from June 1997 to 
August 2006. \citet{c08} present those radial velocity data  
and include a longer time baseline of lower precision 
CORAVEL data to add modest constraints to the stellar binary 
orbit. They model acceptable periods ranging from 40 to 120 years 
for the stellar binary system. Their orbital solution for the 
planet, {HD 196885 Ab}, has a period of 3.69 years, eccentricity 
of 0.462, and velocity semi-amplitude of $40.5 \pm 2.3$ \ms. We cannot 
resolve the inconsistencey between their velocity amplitude and the 
larger velocity amplitude of $53.9 \pm 3.7$ \ms\ that we measure.

\section{Discussion}
  
We report Doppler velocities for six exoplanet discoveries from Lick Observatory.
The planets are all more massive than Jupiter and have significant orbital 
eccentricity with periods ranging from 0.9 to 7.6 years. 
HD30562b has \Msini\ = 1.29 \Mjup, an orbital eccentricity of 0.76,
and an orbital period of 3 years.  HD 86264b has \Msini\ = 7 \Mjup, 
eccentricity of 0.7 and an orbital period of 4 years.  HD 87883b has a mass,
\Msini = 1.78 \Mjup, eccentricity of 0.53 and an orbital period of 7.6 years. 
HD89307b has \Msini\ = 1.78 \Mjup, eccentricity of 0.241 and an orbital period 
of 5.9 years. HD148427b has a mass, \Msini\ = 0.96 \Mjup, a more modest eccentricity 
of 0.16 and an orbital period of 0.9 years. HD196885Ab has \Msini\ = 2.58 \Mjup, 
eccentricity of 0.48 and an orbital period of 3.65 years. 
Among the planets presented in this paper, the high mass 
planets orbit at wider separations and have higher eccentricity 
orbits.

The host stars HD~86264 and HD~148427 are essentially identical in mass 
($M_* \approx 1.4$~\msun) and chemical composition ([Fe/H] $\approx + 0.2$),
but differ in their evolutionary states. 
HD~86264 is a F7V star, while JD~148427 resides on the subgiant branch 
as a K0IV star.  Like most massive main-sequence stars, HD~86264 has 
moderate rotation with \vsini\ = 12.8 \kms and modest chromospheric 
activity. As a result, the Doppler precision is much poorer for HD~86264 than 
for HD~148427 with internal errors of 19\ms\ vs 6.8 \ms, respectively. 
Therefore, it would have been impossible to detect the planet 
found around HD~148427 if it had been orbiting a dwarf star of the same 
mass, like HD~86264. This emphasizes the value in searching for planets
around stars in their cooler, evolved states \citep{j07, s08}. 

Before the detection of planets orbiting other stars, 
it was expected that exoplanets would reside on nearly 
circular orbits, like planets in our own solar system, as 
a result of eccentricity damping in protoplanetary 
disks. However, about one third of Doppler-detected exoplanets have 
measured orbital eccentricities greater than 0.3. As a result, a number of 
mechanisms have been proposed for exciting eccentricity in the orbits of gas 
giant planets (see \citet{fr08} and extensive references therein), including 
perturbations by stellar companions, scattering in the protoplanetary disk, 
resonant interactions between planet embryos and tidal interactions with the disk. 

Planet-planet interactions appear to provide a mechanism that is able 
to reproduce the observed eccentricity distribution. After dissipation 
of the protoplanetary disk, eccentricities can grow rapidly and lead to 
graviatational encounters between planets \citep{cfmr08, fr08, jt08}. 
\citep{fr08} find that simulations of encounters between unequal mass 
planets produce fewer collisions and a broader range of final eccentricities
that reproduce the observed eccentricity distribution.   
\citet{jt08} find that there are many different sets of initial 
conditions that can lead to similar ``relaxed'' eccentricity distributions and 
note that details of initial conditions may therefore be impossible to 
deduce from the final observed states. In order to explain the observed 
eccentricity distribution, \citet{jt08} expect that one or two additional gas giant 
planets must reside in most exoplanet systems.  A natural outcome of 
planet-planet scattering, a significant number of ejected planets and 
non-coplanar systems are expected.

HD196885A is a star with a massive planet in a binary stellar system.
While exoplanets have been found in several binary systems, 
this system is unusual because of the small projected linear separation 
of the stellar components. The current best solution places the M dwarf
companion at a projected angular separation corresponding to 
only 23 AU. One of the challenges for planet formation 
models is the growth of planetesimals from meter-sized objects to kilometer-sized
objects. This challenge is even greater in close binary systems, yet the 
primary star hosts a fairly massive planet with a semi-major axis of 
2.37 AU. A good measurement of the semi-major axis of the stellar binary would
help to understand how this planet could have formed and survived the dynamics
of this challenging environment. 

Most of the host stars presented in this paper 
have high (super-solar) metallicity. Thus, we note that the 
planet-metallicity correlation for gas giant planets \citep{fv05, sa04} 
continues to hold for longer orbital periods.

\acknowledgements
We gratefully acknowledge the dedication and support of the Lick
Observatory staff. DAF acknowledges research support from NASA grant NNX08AF42G.
JAJ is an NSF Astronomy and Astrophysics Postdoctoral Fellow with support 
from the NSF grant AST-0702821. A.W.H gratefully acknowledges support from 
a Townes Postdoctoral Fellowship at the U.C. Berkeley Space Sciences Laboratory.
The authors extend thanks to those of Hawaiian ancestry on whose 
sacred mountain of Mauna Kea we are privileged to be guests.  
Without their kind hospitality, the Keck observations of HD~87883 
would not have been possible. This research has made use of 
the SIMBAD database, operated at CDS, Strasbourg, France, and of 
NASA's Astrophysics Data System Bibliographic Services.

\clearpage

\clearpage

%\begin{figure}
%\plotone{figs/30562.a2.mcmc1.hist.eps}
%\figcaption{MCMC analysis for HD~30562 shows good 
%agreement with the LMMC Keplerian parameters. There is a
%sharp peak in the probability distribution at 1180 days and the orbital parameters
%have reasonably narrow distributions.
%\label{fig_pdf_30562}}
%\end{figure}
%\clearpage

\clearpage

\clearpage

\clearpage

\clearpage

\clearpage

\begin{figure}
\plotone{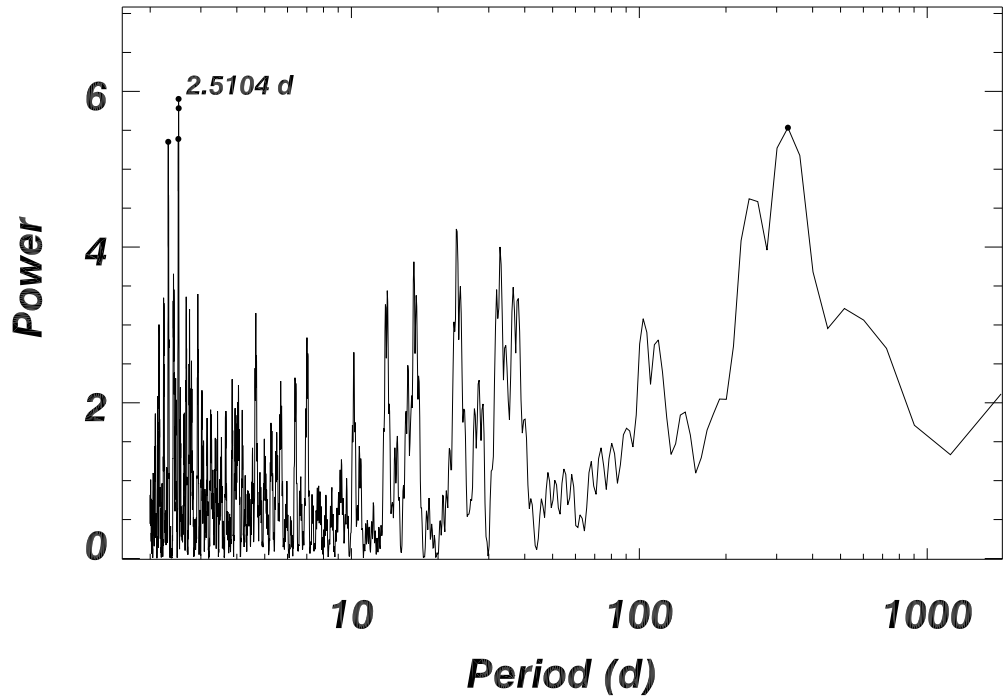}
\figcaption{Periodogram of residuals to the Keplerian 
model of HD 87883 for higher cadence, higher precision Keck 
radial velocities.A modest peak appears at 2.5 days, but we do not consider
this to be significant.  The residual Keck data do not show 
any period with an FAP below 5\%. 
\label{fig_per_87883}} 
\end{figure}
\clearpage

\begin{figure}
\plotone{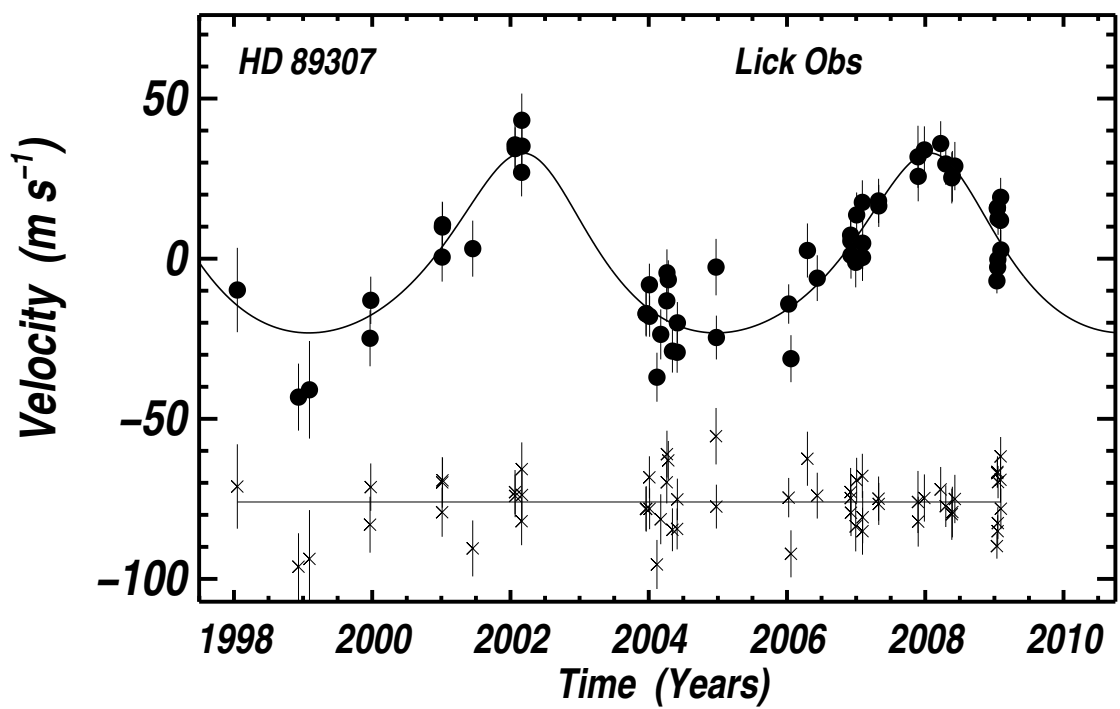}
\figcaption{Radial velocities from Lick Observatory 
are plotted for HD 89307, including 2.8 \ms of jitter 
added in quadrature to the internal errors. 
The best fit Keplerian 
model is overplotted with a solid line and yields 
an orbital period 5.9 years, velocity amplitude 
of 28.9 \ms and eccentricity of 0.24. 
With the assumed stellar mass of 1.028 \msun\ 
we derive a planet mass, \Msini = 1.78 \Mjup\ 
and semi-major axis of 3.27 AU. 
\label{fig_rvfit_89307}}
\end{figure}
\clearpage

\begin{figure}
\plotone{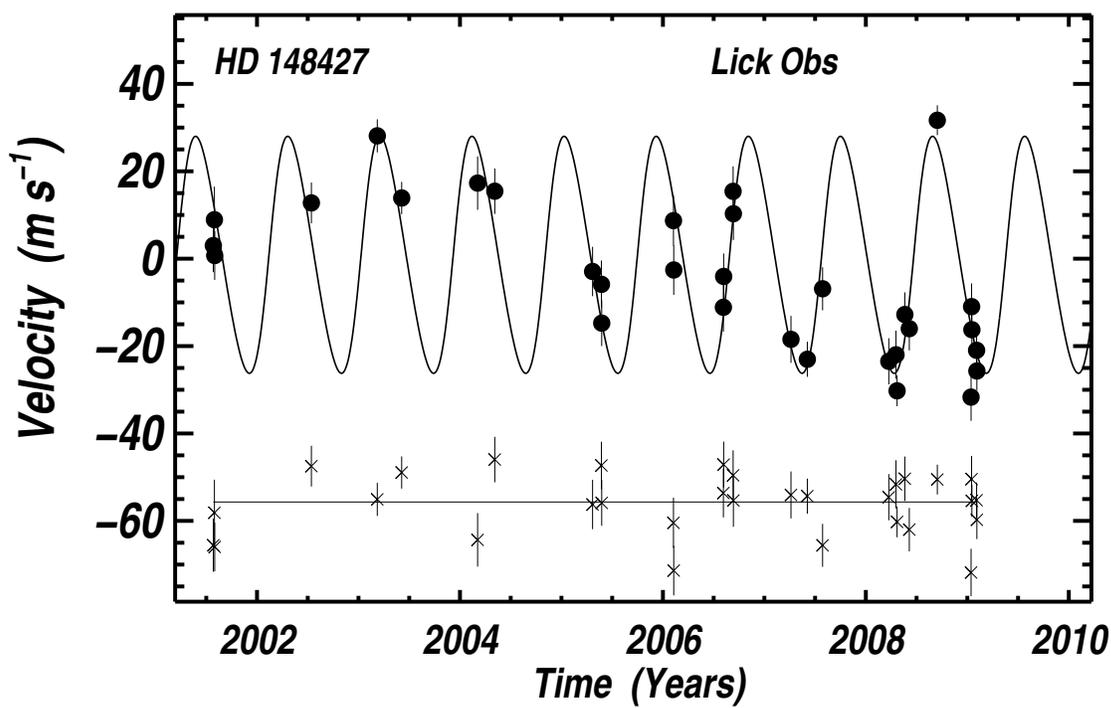}
\figcaption{Time series radial velocities are shown for the 
K0IV star, HD 148427 and include 3.5 \ms of jitter added 
in quadrature with the formal measurement errors. 
The dashed line shows a Keplerian best fit model 
with an orbital period of 0.9 years, velocity amplitude 
of 27.7 \ms and eccentricity of 0.16. 
Adopting a stellar mass of 1.45 \msun we derive a planet mass, 
\Msini = 0.96 \Mjup\ and orbital radius of 0.93 AU.
\label{fig_rvfit_148427}}
\end{figure}
\clearpage

\begin{figure}
\plotone{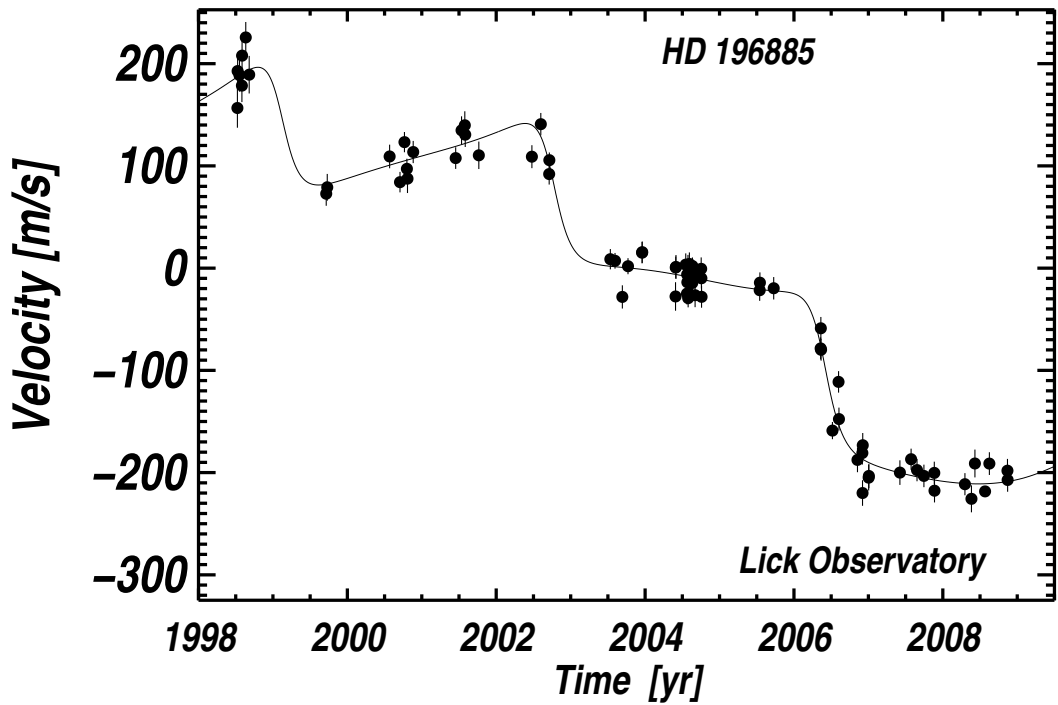}
\figcaption{Time series radial velocities for HD 196885 
are best fit with a double Keplerian model that includes 
the known M dwarf stellar companion and a second planetary 
companion.  An assumed jitter of 2 \ms was added in quadrature
to the formal uncertainties to model this system. 
\label{fig_rvfit_196885}}
\end{figure}
\clearpage

\begin{figure}
\plotone{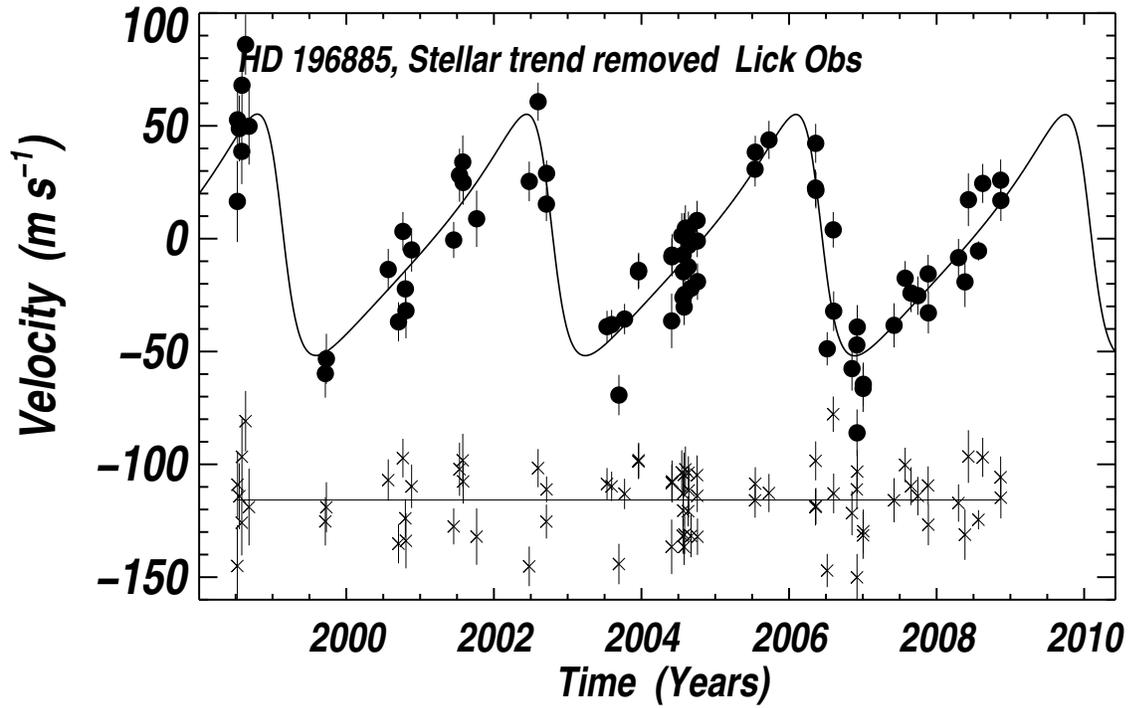}
\figcaption{The Keplerian model for HD 196885Ab is shown with 
the Keplerian model for the stellar companion subtracted off. 
The residual data are best fit by a planetary companion
with \Msini = 2.58 \Mjup\ in an orbit with a semi-major axis of 2.37 AU.  
\label{fig_rvfit_196885b}}
\end{figure}
\clearpage
  
\begin{figure}
\plotone{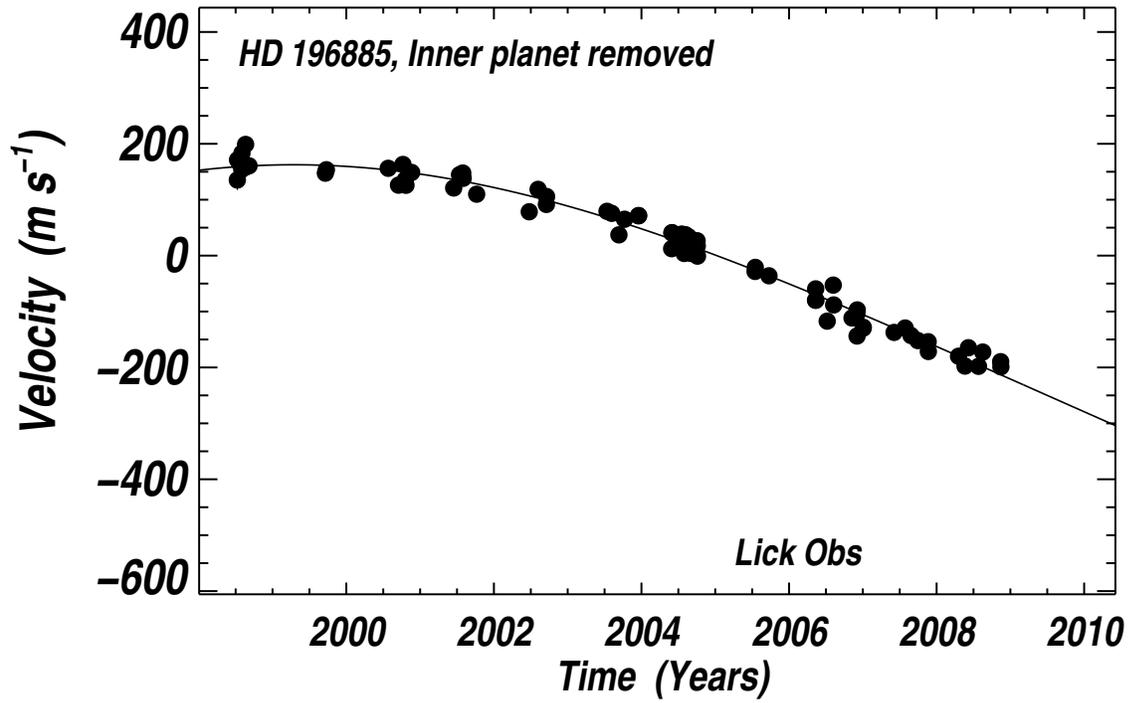}
\figcaption{Residual velocities for HD 196885 A are plotted 
after removing the modeled velocities from the planet. 
The Keplerian model for the stellar companion 
has a fixed period of 120 years, however the phase coverage
is poor and the \rchisq fit is nearly
constant for periods from 40 to 200 years. 
\label{fig_rvfit_196885c}}
\end{figure}
\clearpage

\begin{deluxetable}{lllllll}
\tablecaption{Stellar Parameters\label{tab_t1}}
\rotate
\tablewidth{0pt}
\tablehead{
  \multicolumn{1}{l}{Parameter} & 
  \multicolumn{1}{l}{HD 30562}  & 
  \multicolumn{1}{l}{HD 86264}  & 
  \multicolumn{1}{l}{HD 87883}  &
  \multicolumn{1}{l}{HD 89307}  & 
  \multicolumn{1}{l}{HD 148427} & 
  \multicolumn{1}{l}{HD 196885A} 
}
\startdata
Spectral Type      & F8V              &    F7V           & K0V            & G0V                  & K0IV             &  F8V                 \\
Distance (pc)      & 26.5             &    72.6          & 18.1           & 30.9                 & 59.3             &  33                  \\
\bv                & 0.63             &    0.46          & 0.96           & 0.640                & 0.93             &  0.509               \\
\teff (K)          & 5861 (44)        &    6210 (44)     & 4980 (44)      & 5950 (44)            & 5052 (44)        &  6254 (44)           \\
\logg              & 4.09 (0.10)      &    4.02 (0.10)   & 4.58 (0.10)    & 4.414 (0.10)         & 3.59 (0.10)      &  4.31 (0.10)         \\
\fe                & +0.243 (0.04)    &    +0.202 (0.04) & +0.093 (0.04)  & $-0.14$ (0.04)       & +0.154 (0.04)    &  +0.22 (0.04)        \\
\vsini\ \kms       & 4.9 (0.50)       &    12.8 (0.50)   & 2.17 (0.50)    & 3.21 (0.50)          & 2.13 (0.5)       &  7.8 (0.50)          \\
$V$                & 5.77             &    7.42          & 7.57           & 7.06                 & 6.89             &  6.39                \\
\bc                & -0.064           &    -0.024        & -0.305         & -0.075               & -0.26            &  -0.028              \\
\mv                & 3.65             &    3.10          & 6.27           & 4.57                 & 3.02             &  3.79                \\
\lstar\ (\lsun)    & 2.85 (0.19)      &    4.55 (0.6)    & 0.318 (0.018)  & 1.24 (0.09)          & 6.09 (0.75)      &  2.41 (0.16)         \\
\mstar\ (\msun)    & 1.219 (0.04)     &    1.42 (0.05)   & 0.82 (0.04)    & 1.028 (0.04)         & 1.45 (0.06)      &  1.28 (0.05)         \\
\rstar\ (\rsun)    & 1.637 (0.06)     &    1.88 (0.12)   & 0.76 (0.03)    & 1.05 (0.04)          & 3.22 (0.2)       &  1.31 (0.05)         \\
\shk               & 0.15 (0.005)     &    0.20 (0.017)  & 0.278 (0.021)  & 0.154 (0.008)        & 0.139 (0.017)    &  0.148 (0.018)       \\
\rhk               & -5.064           &   -4.73          & -4.81          & -4.98                & -5.18            &  -5.02               \\
\prot (d)          & 24.2             &    4.15          & 38.6           & 23.7                 & 55.7             &  9.81                \\
Age (Gyr)          & 4.0              &    2.24          & 9.8            & 6.76                 & 2.5              &  3.12                \\
\enddata
\end{deluxetable}
\clearpage
   
\begin{deluxetable}{rrcc}
\tablecaption{Radial Velocities for HD30562\label{tab_t2}}
\tablewidth{0pt}
\tablehead{ 
    \colhead{}   &
    \colhead{RV} & 
    \colhead{$\sigma_{\rm RV}$} \\
       \colhead{JD-2440000} & 
       \colhead{(\ms)} & 
       \colhead{(\ms)} \\ 
 }
\startdata
    11174.91602  &     17.82  &     10.03 \\ 
    11175.80664  &     20.83  &      7.57 \\ 
    11206.75391  &     22.75  &      8.74 \\ 
    11467.97958  &     -9.75  &      6.37 \\ 
    11482.93596  &    -20.05  &      6.73 \\ 
    11534.86523  &    -17.74  &      7.35 \\ 
    11535.85769  &    -14.54  &      5.67 \\ 
    11859.89453  &     -6.57  &      8.25 \\ 
    12937.90033  &     11.27  &      3.48 \\ 
    13282.98926  &     10.04  &      5.10 \\ 
    13341.85063  &     19.69  &      4.90 \\ 
    13360.80337  &     18.80  &      5.44 \\ 
    13389.76837  &     14.19  &      5.10 \\ 
    13391.77977  &     18.54  &      5.32 \\ 
    13743.72993  &    -36.48  &      6.28 \\ 
    13751.78702  &    -34.28  &      5.69 \\ 
    13988.95190  &    -22.96  &      5.50 \\ 
    14049.94301  &    -11.77  &      5.02 \\ 
    14071.95702  &     -0.83  &      6.80 \\ 
    14072.80531  &    -13.94  &      5.34 \\ 
    14073.86159  &    -17.66  &      5.54 \\ 
    14099.84141  &    -19.94  &      5.19 \\ 
    14103.77660  &    -10.93  &      4.77 \\ 
    14337.98200  &     -1.27  &      4.99 \\ 
    14374.02309  &      2.84  &      5.64 \\ 
    14428.95087  &     10.40  &      5.07 \\ 
    14447.83085  &      3.26  &      3.81 \\ 
    14461.87088  &      4.07  &      5.23 \\ 
    14517.65772  &      9.41  &      7.29 \\ 
    14548.65551  &     21.61  &      5.11 \\ 
    14723.02289  &     35.82  &      3.85 \\ 
    14724.01284  &     37.37  &      3.55 \\ 
    14756.94272  &     18.69  &      4.43 \\ 
    14783.85702  &    -29.31  &      5.30 \\ 
    14784.89422  &    -22.55  &      6.08 \\ 
    14785.94216  &    -22.72  &      4.06 \\ 
    14845.77865  &    -27.85  &      3.24 \\ 
    14846.75067  &    -23.42  &      4.61 \\ 
    14847.73611  &    -32.97  &      4.38 \\ 
    14848.75761  &    -37.29  &      3.70 \\ 
    14849.72956  &    -37.37  &      3.92 \\ 
    14850.75859  &    -21.96  &      4.11 \\ 
    14863.74239  &    -18.96  &      4.57 \\ 
    14864.68794  &    -17.42  &      5.47 \\ 
    14865.70517  &    -16.68  &      4.62 \\ 
\enddata
\end{deluxetable}
\clearpage

\begin{deluxetable}{llllllll}
\tablecaption{Orbital Parameters\label{tab_t3}}
\rotate
\tablewidth{0pt}
\tablehead{
  \multicolumn{1}{l}{Parameter}   & 
  \multicolumn{1}{l}{HD 30562b}   & 
  \multicolumn{1}{l}{HD 86264b}   & 
  \multicolumn{1}{l}{HD 87883b}   &
  \multicolumn{1}{l}{HD 89307b}   & 
  \multicolumn{1}{l}{HD 148427b}  & 
  \multicolumn{1}{l}{HD 196885Ab} 
} 
\startdata
P (d)            &   1157 (27)    & 1475 (55)   &  2754 (87)     &  2157 (63)    &  331.5 (3.0)    &  1333 (15)     \\
$K$ (\ms)        &   33.7 (2.2)   & 132 (33)    &  34.7 (4.5)    &  28.9 (2.2)   &  27.7 (2)       &  53.9 (3.7)    \\
$e$              &   0.76 (0.05)  & 0.7 (0.2)   &  0.53 (0.12)   &  0.241 (0.07) &  0.16 (0.08)    &  0.48 (0.06)   \\
\Tp\ (JD)        &  10131.5 (14)  & 15172 (114) &  11139 (90)    &  10228 (80)   &  13991 (15)     &  11992 (12)    \\
$\omega$ (deg)   &   81 (10)      & 306 (10)    &  291 (15)      &  36 (52)      &  277 (68)       &  78 (7.6)      \\
Trend (\msyr)    &   \nodata      & 0.005       &  \nodata       &  \nodata      &  \nodata        &  \nodata       \\
$a$ (AU)         &   2.3 (0.02)   & 2.86 (0.07) &  3.6 (0.08)    &  3.27 (0.07)  &  0.93 (0.01)    &  2.37 (0.02)   \\
\Msini (\Mjup)   &   1.29 (0.08)  & 7.0 (1.6)   &  1.78 (0.34)   &  1.78 (0.13)  &  0.96 (0.1)     &  2.58 (0.16)   \\
$N_{\rm obs}$     &   45           & 37          &  69            &  59           &  31             &  76            \\
jitter \ms       &   2.9          &  3.3        &  4.5           &  2.8          &  3.5            &  2.0           \\
RMS (\ms)        &   7.58          & 26.2       &  9.2           &  9.9          &  7.0            &  14.7          \\
\rchisq          &   1.31         & 1.21        &  1.71          &  1.37         &  1.08           &  1.58          \\
FAP              &   $< 0.0001$   & $<0.0001$   &  $<0.0001$     &  $<0.0001$    &  $<0.0001$      &  $<0.0001$     \\  
\enddata
\end{deluxetable}                
\clearpage

\begin{deluxetable}{rrcc}
\tablecaption{Radial Velocities for HD86264\label{tab_t4}}
\tablewidth{0pt}
\tablehead{ 
    \colhead{}   &
    \colhead{RV} & 
    \colhead{$\sigma_{\rm RV}$} \\
       \colhead{JD-2440000} & 
       \colhead{(\ms)} & 
       \colhead{(\ms)} \\ 
 }
\startdata
    11913.97949  &   -124.29  &     21.75 \\ 
    11914.95215  &   -106.63  &     25.94 \\ 
    11915.97266  &   -133.09  &     22.51 \\ 
    11927.95800  &   -114.71  &     20.32 \\ 
    11946.87109  &   -110.60  &     19.18 \\ 
    12298.95410  &    124.05  &     19.34 \\ 
    12333.84668  &    101.74  &     20.59 \\ 
    13033.93945  &    -58.99  &     19.57 \\ 
    13101.71289  &    -51.24  &     24.08 \\ 
    13389.83370  &    -48.75  &     18.64 \\ 
    13391.97233  &    -96.12  &     18.67 \\ 
    13392.81923  &    -45.12  &     14.91 \\ 
    13756.87898  &    133.09  &     19.13 \\ 
    13869.71119  &     94.29  &     16.79 \\ 
    13895.69225  &     62.54  &     27.79 \\ 
    14073.04851  &     13.04  &     17.77 \\ 
    14074.02220  &    -30.92  &     18.04 \\ 
    14099.03887  &    -18.92  &     24.35 \\ 
    14099.99183  &     -2.54  &     19.56 \\ 
    14103.92013  &    -11.77  &     18.46 \\ 
    14133.86964  &    -21.51  &     10.55 \\ 
    14135.79582  &     11.93  &     13.57 \\ 
    14165.79559  &     91.33  &     17.90 \\ 
    14461.93014  &    -64.70  &     24.53 \\ 
    14548.77584  &    -89.39  &     16.58 \\ 
    14574.75097  &    -40.01  &     16.78 \\ 
    14578.76834  &    -26.36  &     14.60 \\ 
    14622.69367  &    -59.33  &     31.96 \\ 
    14844.97356  &    -87.15  &     12.52 \\ 
    14846.05428  &    -91.75  &     15.39 \\ 
    14846.97546  &   -114.10  &     13.59 \\ 
    14848.92699  &    -67.28  &     12.89 \\ 
    14849.87817  &   -105.34  &     12.34 \\ 
    14850.92234  &    -80.79  &     13.45 \\ 
    14863.88580  &    -93.69  &     16.25 \\ 
    14864.89708  &   -132.25  &     20.60 \\ 
    14865.90181  &    -78.03  &     15.50 \\ 
\enddata
\end{deluxetable}
\clearpage

\begin{deluxetable}{rrcc}
\tablecaption{Radial Velocities for HD87883\label{tab_t5}}
\tablewidth{0pt}
\tablehead{ 
    \colhead{}   &
    \colhead{RV} & 
    \colhead{$\sigma_{\rm RV}$} &
    \colhead{} \\ 
       \colhead{JD-2440000} & 
       \colhead{(\ms)} & 
       \colhead{(\ms)} & 
       \colhead{Observatory}  \\
 }
\startdata
     1998.93369  &     12.80  &      7.97  &    L   \\ 
     1999.17408  &     33.63  &      7.17  &    L   \\ 
     1999.97128  &     18.35  &      6.44  &    L   \\ 
     2001.01165  &      0.88  &      5.85  &    L   \\ 
     2001.01429  &     -5.14  &      6.10  &    L   \\ 
     2001.01709  &      5.80  &      5.87  &    L   \\ 
     2001.04992  &     -1.50  &      6.07  &    L   \\ 
     2001.09908  &     13.53  &      6.08  &    L   \\ 
     2001.10167  &     13.45  &      6.32  &    L   \\ 
     2001.45430  &    -17.65  &      6.10  &    L   \\ 
     2002.06567  &    -14.97  &      4.72  &    L   \\ 
     2002.16120  &      5.50  &      5.98  &    L   \\ 
     2002.16649  &     -9.29  &      6.11  &    L   \\ 
     2003.13049  &    -22.78  &      4.23  &    L   \\ 
     2003.96032  &    -20.66  &      4.64  &    L   \\ 
     2003.96316  &    -18.05  &      5.25  &    L   \\ 
     2004.00956  &    -18.31  &      4.81  &    L   \\ 
     2004.26352  &    -42.47  &      5.83  &    L   \\ 
     2005.05229  &    -41.12  &      5.03  &    L   \\ 
     2005.05815  &    -32.01  &      4.92  &    L   \\ 
     2006.92539  &     20.38  &      5.46  &    L   \\ 
     2006.95556  &     42.47  &      2.80  &    K   \\ 
     2007.00210  &     28.05  &      5.22  &    L   \\ 
     2007.00472  &     36.03  &      5.33  &    L   \\ 
     2007.00740  &     26.61  &      5.16  &    L   \\ 
     2007.09195  &     18.36  &      5.43  &    L   \\ 
     2007.17675  &     19.74  &      5.27  &    L   \\ 
     2007.26151  &     33.31  &      4.83  &    L   \\ 
     2007.98755  &     14.49  &      5.27  &    L   \\ 
     2008.14057  &     16.07  &      4.63  &    L   \\ 
     2008.22279  &      9.73  &      4.87  &    L   \\ 
     2008.29645  &      0.97  &      5.21  &    L   \\ 
     2008.38397  &     -4.94  &      6.73  &    L   \\ 
     2008.39214  &     20.11  &      5.62  &    L   \\ 
     2008.42769  &      2.86  &      6.03  &    L   \\ 
     2008.47985  &      1.75  &      4.16  &    K   \\ 
     2008.85326  &     17.75  &      4.20  &    K   \\ 
     2008.85584  &     10.71  &      4.01  &    K   \\ 
     2008.85868  &     -6.52  &      3.96  &    K   \\ 
     2008.87491  &    -10.34  &      8.02  &    L   \\ 
     2008.88886  &      3.65  &      3.88  &    K   \\ 
     2008.92978  &     17.89  &      3.96  &    K   \\ 
     2008.93511  &     11.34  &      5.32  &    K   \\ 
     2008.93805  &     -2.68  &      3.92  &    K   \\ 
     2008.94051  &      5.14  &      4.05  &    K   \\ 
     2008.94338  &     10.12  &      3.93  &    K   \\ 
     2009.02007  &     -4.52  &      4.02  &    K   \\ 
     2009.03607  &      7.99  &      2.81  &    L   \\ 
     2009.03925  &     15.00  &      3.49  &    L   \\ 
     2009.04198  &      1.03  &      4.21  &    K   \\ 
     2009.04426  &     -0.41  &      3.77  &    L   \\ 
     2009.04697  &     -2.47  &      3.67  &    L   \\ 
     2009.04970  &    -13.64  &      3.63  &    L   \\ 
     2009.05243  &    -11.98  &      4.53  &    L   \\ 
     2009.08818  &     -3.23  &      4.63  &    L   \\ 
     2009.09078  &    -11.39  &      6.23  &    L   \\ 
     2009.09116  &    -21.38  &      4.11  &    K   \\ 
     2009.09359  &     10.95  &      6.34  &    L   \\ 
     2009.09890  &    -12.31  &      4.24  &    K   \\ 
     2009.26318  &     -7.77  &      4.14  &    K   \\ 
     2009.26863  &    -14.77  &      4.20  &    K   \\ 
     2009.28229  &    -15.93  &      4.35  &    K   \\ 
     2009.33710  &    -10.03  &      3.82  &    K   \\ 
     2009.36175  &     -3.13  &      3.99  &    K   \\ 
     2009.41635  &     -4.02  &      3.86  &    K   \\ 
     2009.41906  &     -9.49  &      3.92  &    K   \\ 
     2009.42181  &    -15.32  &      4.16  &    K   \\ 
     2009.42461  &     -1.15  &      3.94  &    K   \\ 
     2009.42729  &    -13.37  &      4.31  &    K   \\ 
\enddata
\end{deluxetable}
\clearpage

\begin{deluxetable}{rrr}
\tablecaption{Radial Velocities for HD89307\label{tab_t6}}
\tablewidth{0pt}
\tablehead{ 
    \colhead{}   &
    \colhead{RV} & 
    \colhead{$\sigma_{\rm RV}$} \\
       \colhead{JD-2440000} & 
       \colhead{(\ms)} & 
       \colhead{(\ms)} \\ 
 }
\startdata
    10831.87988  &     -9.78  &     12.67 \\ 
    11155.06152  &    -43.23  &      9.82 \\ 
    11212.96387  &    -40.94  &     14.81 \\ 
    11533.03711  &    -24.86  &      7.96 \\ 
    11536.94238  &    -13.00  &      6.46 \\ 
    11914.03613  &      0.51  &      6.77 \\ 
    11914.97949  &      9.90  &      7.03 \\ 
    11916.00391  &     10.64  &      6.00 \\ 
    12075.69141  &      3.13  &      7.97 \\ 
    12299.01562  &     35.54  &      5.90 \\ 
    12299.81152  &     34.38  &      5.56 \\ 
    12333.89355  &     26.96  &      6.57 \\ 
    12334.78516  &     43.23  &      7.53 \\ 
    12335.84668  &     35.15  &      7.04 \\ 
    12991.03027  &    -17.15  &      5.84 \\ 
    12992.05762  &    -17.21  &      6.13 \\ 
    13009.01367  &     -8.14  &      5.50 \\ 
    13009.97070  &    -17.99  &      5.28 \\ 
    13048.88880  &    -37.26  &      7.21 \\ 
    13068.85645  &    -23.63  &      6.92 \\ 
    13100.75994  &    -13.11  &      6.25 \\ 
    13101.78418  &     -4.40  &      6.38 \\ 
    13108.80719  &     -6.78  &      5.50 \\ 
    13130.74121  &    -28.86  &      5.65 \\ 
    13155.68164  &    -29.22  &      5.29 \\ 
    13156.69141  &    -20.03  &      5.39 \\ 
    13361.03331  &     -2.62  &      8.45 \\ 
    13362.97417  &    -24.75  &      6.35 \\ 
    13744.96252  &    -14.16  &      4.98 \\ 
    13756.89883  &    -31.22  &      6.39 \\ 
    13843.85424  &      2.56  &      7.64 \\ 
    13895.70819  &     -6.08  &      6.18 \\ 
    14072.05486  &      7.35  &      6.63 \\ 
    14073.05723  &      5.55  &      5.96 \\ 
    14074.03192  &      1.07  &      6.34 \\ 
    14099.05421  &     -1.15  &      6.93 \\ 
    14103.99258  &     13.67  &      6.01 \\ 
    14133.92363  &     17.59  &      5.87 \\ 
    14134.86984  &      0.39  &      6.40 \\ 
    14135.81388  &      4.83  &      7.31 \\ 
    14219.75881  &     18.05  &      5.90 \\ 
    14220.77130  &     16.48  &      5.49 \\ 
    14428.07576  &     31.82  &      9.01 \\ 
    14429.07464  &     25.71  &      6.91 \\ 
    14461.97798  &     33.97  &      6.42 \\ 
    14547.88772  &     35.95  &      5.99 \\ 
    14574.81520  &     29.59  &      5.42 \\ 
    14606.73102  &     25.25  &      7.14 \\ 
    14609.71447  &     25.74  &      7.05 \\ 
    14622.73115  &     28.91  &      6.63 \\ 
    14844.94877  &     -6.94  &      3.47 \\ 
    14846.99865  &     15.72  &      3.79 \\ 
    14847.99892  &     15.93  &      4.15 \\ 
    14848.94512  &     -2.65  &      4.31 \\ 
    14849.93880  &     -0.10  &      4.24 \\ 
    14850.94371  &     12.47  &      4.37 \\ 
    14863.93069  &     12.09  &      6.65 \\ 
    14864.88307  &     19.22  &      5.49 \\ 
    14865.91973  &      2.72  &      5.84 \\ 
\enddata
\end{deluxetable}
\clearpage

\begin{deluxetable}{rrr}
\tablecaption{Radial Velocities for HD148427\label{tab_t7}}
\tablewidth{0pt}
\tablehead{ 
    \colhead{}   &
    \colhead{RV} & 
    \colhead{$\sigma_{\rm RV}$} \\
       \colhead{JD-2440000} & 
       \colhead{(\ms)} & 
       \colhead{(\ms)} \\ 
 }
\startdata
    12117.77468  &      3.02  &      5.84 \\ 
    12121.73047  &      8.93  &      7.15 \\ 
    12122.74609  &      0.73  &      4.99 \\ 
    12470.75977  &     12.77  &      3.94 \\ 
    12707.99301  &     28.24  &      3.50 \\ 
    12795.82135  &     13.79  &      3.39 \\ 
    13069.05859  &     17.30  &      5.59 \\ 
    13130.97559  &     15.45  &      4.56 \\ 
    13482.91224  &     -2.92  &      5.06 \\ 
    13514.81401  &     -5.86  &      4.83 \\ 
    13515.77733  &    -14.75  &      4.61 \\ 
    13774.07313  &      8.72  &      5.17 \\ 
    13775.06466  &     -2.58  &      5.11 \\ 
    13953.67145  &    -11.12  &      4.99 \\ 
    13954.67293  &     -4.05  &      4.61 \\ 
    13988.64826  &     15.42  &      5.15 \\ 
    13989.64802  &     10.30  &      5.48 \\ 
    14196.94310  &    -18.46  &      4.76 \\ 
    14255.88785  &    -23.03  &      3.60 \\ 
    14310.74223  &     -6.88  &      4.23 \\ 
    14549.03808  &    -23.48  &      4.66 \\ 
    14574.91787  &    -21.97  &      4.91 \\ 
    14578.89804  &    -30.23  &      3.05 \\ 
    14606.89543  &    -12.78  &      4.42 \\ 
    14622.86542  &    -16.01  &      4.29 \\ 
    14723.66497  &     31.67  &      2.92 \\ 
    14845.09173  &    -31.67  &      5.13 \\ 
    14847.10053  &    -10.97  &      4.69 \\ 
    14848.09531  &    -16.06  &      3.20 \\ 
    14865.07943  &    -20.97  &      3.61 \\ 
    14866.07445  &    -25.72  &      4.03 \\ 
\enddata
\end{deluxetable}
\clearpage

\begin{deluxetable}{rrr}
\tablecaption{Radial Velocities for HD196885A\label{tab_t8}}
\tablewidth{0pt}
\tablehead{ \colhead{JD} & \colhead{RV} & \colhead{Uncertainties}   \\
  \colhead{-2440000.} & \colhead{(\ms)} & \colhead{(\ms)}   \\}
\startdata
    11004.87786  &    156.62  &     17.99   \\ 
    11005.91113  &    192.76  &     14.54   \\ 
    11014.87598  &    188.81  &     14.59   \\ 
    11026.84668  &    178.52  &     14.52   \\ 
    11027.87109  &    207.85  &     16.91   \\ 
    11045.85156  &    225.65  &     13.38   \\ 
    11062.77637  &    189.21  &     17.01   \\ 
    11440.64692  &     72.65  &     10.70   \\ 
    11445.72070  &     79.05  &     11.09   \\ 
    11751.85840  &    109.24  &      9.18   \\ 
    11802.79239  &     84.09  &      8.69   \\ 
    11824.63919  &    123.29  &      8.56   \\ 
    11836.69527  &     97.32  &      8.54   \\ 
    11839.62402  &     87.54  &     12.24   \\ 
    11867.63857  &    113.28  &      9.64   \\ 
    12075.96484  &    107.64  &      7.97   \\ 
    12104.88281  &    134.78  &     11.78   \\ 
    12120.84375  &    139.73  &     11.79   \\ 
    12122.85938  &    130.53  &      9.83   \\ 
    12189.63457  &    110.52  &     12.50   \\ 
    12449.96387  &    109.00  &      8.79   \\ 
    12493.81152  &    140.81  &      8.45   \\ 
    12534.77637  &     92.01  &      7.53   \\ 
    12535.67998  &    105.53  &      5.70   \\ 
    12834.92419  &      8.73  &      7.03   \\ 
    12857.93443  &      7.09  &      6.51   \\ 
    12893.79688  &    -28.15  &      8.96   \\ 
    12921.69336  &      2.33  &      6.73   \\ 
    12990.64258  &     15.11  &      7.78   \\ 
    12991.59863  &     15.60  &      7.85   \\ 
    13154.95898  &    -27.71  &     12.08   \\ 
    13155.92090  &      0.59  &      8.15   \\ 
    13156.95117  &      1.17  &      9.31   \\ 
    13204.88432  &      3.41  &      9.82   \\ 
    13210.93750  &    -24.94  &      7.91   \\ 
    13211.91406  &     -6.21  &      7.52   \\ 
    13212.85449  &    -13.81  &      6.89   \\ 
    13215.90261  &      1.96  &      9.82   \\ 
    13216.86158  &    -30.02  &      8.00   \\ 
    13218.85502  &    -24.73  &      8.53   \\ 
    13222.87458  &      4.10  &     10.15   \\ 
    13236.80479  &    -14.96  &      6.69   \\ 
    13237.80842  &      2.00  &      7.35   \\ 
    13239.87256  &     -5.86  &     10.66   \\ 
    13250.83496  &    -26.50  &      9.40   \\ 
    13280.76758  &     -0.69  &      8.72   \\ 
    13281.72754  &     -9.92  &      7.17   \\ 
    13282.69531  &    -28.07  &      8.07   \\ 
    13567.90576  &    -21.60  &      7.69   \\ 
    13568.94066  &    -14.24  &      7.28   \\ 
    13636.69163  &    -19.68  &      8.45   \\ 
    13866.95528  &    -78.52  &      8.18   \\ 
    13867.88733  &    -58.85  &      8.63   \\ 
    13868.98856  &    -79.86  &      8.03   \\ 
    13924.88196  &   -159.16  &      7.36   \\ 
    13954.79810  &   -111.27  &      7.85   \\ 
    13956.86757  &   -147.65  &      8.80   \\ 
    14047.64309  &   -187.58  &      9.78   \\ 
    14071.60554  &   -180.85  &      8.84   \\ 
    14072.60604  &   -219.97  &     10.38   \\ 
    14073.57926  &   -173.23  &      9.71   \\ 
    14102.59081  &   -204.95  &     10.44   \\ 
    14103.58541  &   -203.28  &      9.64   \\ 
    14255.98494  &   -199.99  &      9.78   \\ 
    14310.86948  &   -186.92  &      7.63   \\ 
    14339.83862  &   -197.52  &      8.41   \\ 
    14373.72263  &   -203.25  &      8.51   \\ 
    14424.65135  &   -200.30  &      8.46   \\ 
    14425.58582  &   -217.71  &      9.11   \\ 
    14575.00470  &   -211.34  &      8.28   \\ 
    14606.98432  &   -225.65  &     11.16   \\ 
    14623.95213  &   -191.09  &     11.68   \\ 
    14673.88099  &   -219.00  &      4.25   \\ 
    14694.90059  &   -191.19  &      8.66   \\ 
    14783.63365  &   -198.13  &      9.19   \\ 
    14784.60365  &   -207.20  &      9.13   \\ 
\enddata
\end{deluxetable}
\clearpage


\clearpage 
\begin{thebibliography}{} 
\parsep 0pt 

\bibitem[Baliunas et al.(1995)]{b95}  
Baliunas, S.~L., et al.\ 1995, ApJ, 438, 269

\bibitem[Baliunas et al.(1997)]{b97}
Baliunas, S.~L., Henry, G.~W., Donahue, R.~A., Fekel, F.~C. \& Soon, W.~H.\
1997, ApJ, 474, L119

\bibitem[Bouchy et al.(2009)]{bou09}
Bouchy, F., Mayor, M., Lovis, C., Udry, S., Benz, W., Bertaux, J.-L.,
Delfosse, X., Mordasini, C., Pepe, F., Queloz, D. \& Segransan, D., 
\aap, 496, 572

\bibitem[Butler et al.(1996)]{bu96} 
Butler, R.~P., Marcy, 
G.~W., Williams, E., McCarthy, C., Dosanjh, P.
\& Vogt, S.~S.\ 1996, \pasp, 108, 500 

\bibitem[Butler et al.(2006)]{bu06}
Butler, R. P., Wright, J. T., Marcy, G. W., Fischer, D. A., 
Vogt, S. S., Tinney, C. G., Jones, H. R. A., Carter, B.D., 
Johnson, J. A., McCarthy, C., Penny, A. J. 
2006, \apj, 646, 505

\bibitem[Chatterjee et al.(2008)]{cfmr08}
Chatterjee, S., Ford, E. B., Matsumura, S., Rasio, F. A. 
2008, \apj, 686, 580

\bibitem[Chauvin et al.(2006)]{ch06} 
Chauvin, G., Lagrange, A.-M., Udry, S., Fusco, T., 
Galland, F., Naef, D., Beuzit, J.-L. \& Mayor, M.\ 2006, \aap, 456, 1165 

\bibitem[Chauvin et al.(2007)]{ch07} 
Chauvin, G., Lagrange, A.-M., Udry, S. \& Mayor, M.\ 2007, \aap, 475, 723 

\bibitem[Correia et al.(2008)]{c08} 
Correia, A.~C.~M., et al.\ 2008, \aap, 479, 271 

\bibitem[Demarque et al.(2004)]{d04}
Demarque, P., Woo., J.-H., Kim, Y.-C. \& Yi, S.~K.\ 2004, \apjs, 155, 667  

\bibitem[Duncan et al.(1991)]{d91}
Duncan, D. K., et al. 1991, \apjs, 76, 383

\bibitem[Driscoll, Fischer, \& Ford(2009)]{dff09}
Driscoll, P., Fischer, D. \& Ford, E.\ 2009, \apj, in press

\bibitem[ESA(1997)]{esa97}
ESA 1997, The Hipparcos and Tycho Catalogs. 
ESA-SP 1200 

\bibitem[Fischer \& Valenti(2005)]{fv05}
Fischer, D. A. \& Valenti, J. A. 2005, 
ApJ 622, 1102

\bibitem[Fischer et al.(1999)]{fi99}
Fischer, D. A., Marcy, G.W., Butler, R. P., Vogt, S.S. \& Apps, K.
1999, \pasp 111, 50 

\bibitem[Ford(2005)]{f05} 
Ford, E.~B.\ 2005, \aj, 129, 1706 

\bibitem[Ford \& Rasio (2008)]{fr08}
Ford, E. B., Rasio \& Rasio, F. A. 2008, \apj, 686, 621

\bibitem[Giguere \& Fischer(2009)]{gf09}
Giguere, M. \& Fischer, D.A. 2009, \apj (in prep)

\bibitem[Henry(1999)]{h1999}
Henry, G. W. 1999, \pasp, 111, 845

\bibitem[Henry, Fekel, \& Henry(2007)]{hfh2007}
Henry, G. W., Fekel, F. C., \& Henry, S. M. 2007, \aj, 133, 1421  

\bibitem[Howard et al.(2009)]{how09}
Howard, A. W., Johnson, J. A., Marcy, G. W., Fischer, D. A., 
Wright, J. T., Henry, G. W., Giguere, M. J., Isaacson, H., Valenti, J. A.,
Anderson, J. \& Piskunov, N. E. 2009, \apj in press 

\bibitem[Isaacson \& Fischer(2009)]{if09}
Isaacson, H. \& Fischer, D.A.\ 2009, \apj\ in prep

\bibitem[Johnson et al.(2007)]{j07}
Johnson, J.~A., Fischer,  D.~A., Marcy, G.~W., Wright, J.~T., 
Driscoll, P., Butler, R.~P., Hekker, S., Reffert, S., Vogt, S.~S. 2007,
\apj, 665, 785
 
\bibitem[Juric \& Tremaine(2008)]{jt08}
Juri\'c, M., Tremaine, S. 2008, \apj, 686, 603 

\bibitem[Marcy \& Butler(1992)]{ma92}
Marcy, G.~W. \& Butler, R.~P.\ 1992, \pasp, 104, 270 

\bibitem[Marcy et al.(1997)]{ma97} Marcy, G.~W., Butler, 
R.~P., Williams, E., Bildsten, L., Graham, J.~R., Ghez, A.~M.
\& Jernigan, J.~G.\ 1997, \apj, 481, 926 

\bibitem[Marcy et al.(2002)]{ma02}
Marcy, G., Butler, R. P., Fischer, D., Laughlin, G., Vogt, S.,
Henry, G. \& Pourbaix, D. 2002, \apj, 581, 1375

\bibitem[Mayor et al.(2009a)]{may09a}
Mayor, M., Bonfils, X., Forveille, T., Delfosse, X., Udry, S., Bertaux, J. -L., 
Beust, H., Bouchy, F., Lovis, C., Pepe, F., Perrier, C., Queloz, D., Santos, N. C.
2009, arXiv0906.2780 

\bibitem[Mayor et al.(2009b)]{may09b}
Mayor, M., Udry, S., Lovis, C., Pepe, F., Queloz, D., 
Benz, W., Bertaux, J.-L., Bouchy, F., Mordasini, C., Segransan, D. 2009, 
\aa, 493, 639 

\bibitem[Moutou et al.(2009)]{mou09}
Moutou, C., Mayor, M., Lo Curto, G., Udry, S., Bouchy, F., Benz, W., 
Lovic, C., Naef, D., Pepe, F., Queloz, D. \& Santos, N.-C.
2009, \aap, 496, 513

\bibitem[Noyes et al.(1984)]{noyes84} 
Noyes, R.~W., Hartmann, L., Baliunas, S.~L.,
Duncan, D.~K. \& Vaughan, A.~H.\ 1984, \apj, 279, 763 

\bibitem[Queloz et al.(2001)]{q01}
Queloz, D., Henry, G. W., Sivan, J. P., Baliunas, S. L., 
Beuzit, J. L., Donahue, R. A., Mayor, M., Naef, D., Perrier, C. \&
Udry, S. 2001, \aap, 379, 279

\bibitem[Saar, Butler, Marcy(1998)]{sbm98}
Saar, S. H., Butler, R. P. \& Marcy, G. W. 1998, \apjl, 498, 153

\bibitem[Saar \& Fischer(1999)]{sf00}
Saar, S. H. \& Fischer, D. A. 2000, \apjl, 534, 105

\bibitem[Santos, Israelian \& Mayor(2004)]{sa04}
Santos, N. C., Israelian, G., Mayor, M. 2004, \aap, 415, 1153

\bibitem[Sato et al.(2008)]{s08}
Sato, B., Izumiura, H., Toyota, E., Kambe, E., Ikoma, M., Omiya, M, 
Masuda, S., Takeda, Y., Murata, D., Itoh, Y., Ando, H., Yoshida, M., 
Kokubo, E.,  \& Ida, S. 2008, \pasj, 60, 539

\bibitem[Schneider(2009)]{sch09} 
Schneider, J. 2009, http://exoplanet.eu/catalog.php

\bibitem[Takeda et al.(2007)]{t07} 
Takeda, G., Ford, E.~B., 
Sills, A., Rasio, F.~A., Fischer, D.~A., 
\& Valenti, J.~A.\ 2007, \apjs, 168, 297 

\bibitem[Valenti et. al.(2009)]{v09}
Valenti, J.~A., Fischer, D.~A., Marcy, G.~W., Johnson, J.~A., 
Henry, G.~W., Wright, J.~T., Howard, A.~W., Giguere, M. \& Isaacson, H.\ 
2009, \apj, in press

\bibitem[Valenti \& Fischer(2005)]{vf05} 
Valenti, J.~A. \& Fischer, D.~A.\ 2005, 
\apjs, 159, 141 

\bibitem[Valenti \& Piskunov(1996)]{vp96}
Valenti, J.~A. \& Piskunov, N.\ 1996, \aap, 118, 595
  
\bibitem[Vogt et al.(1994)]{v94}
Vogt, S. S. \etal\ 1994,
SPIE, 2198, 362. 

\bibitem[Vogt et al.(1987)]{v87}
Vogt, S. S. 1987, \pasp, 99, 1214

\bibitem[VandenBerg \& Clem(2003)]{vc03}
VandenBerg, D. A. \& Clem, J. L. 2003, AJ, 126, 778

\bibitem[Wright(2005)]{w05} 
Wright, J. T. 2005,
PASP, 117, 657
  
\bibitem[Wright \& Howard(2009)]{wh09}
Wright, J. T. \& Howard, A. 2009, \apj (in press)

\end{thebibliography}
\end{document}